\documentclass[aps, twocolumn, prb,
showpacs,preprintnumbers,amsmath,amssymb,superscriptaddress]{revtex4-1}
\usepackage{epsfig}
\usepackage{graphics}
\usepackage{subcaption}
\usepackage{xcolor}
\usepackage{amsmath,amssymb,bbold,bm}
\usepackage{mathdots}
\usepackage{mathtools}
\usepackage[font=small,justification=justified]{caption}
\usepackage{cancel}
\usepackage{comment}
\usepackage{currfile}
\usepackage{simplewick}
\usepackage{subcaption}

\newcommand{\tr}{{\rm Tr}}
\newcommand{\bb}{\mathbb}
\newcommand{\eps}{\epsilon}

\newcommand{\bw}[1]{\begin{widetext}}
\newcommand{\ew}[1]{\end{widetext}}
\newcommand{\pmat}[1]{\begin{pmatrix}#1\end{pmatrix}}

\newcommand{\gtappr}{{{\lower4pt\hbox{$>$} } \atop \widetilde{ \ \ \ }}}
\newcommand{\dg}{^{\dagger}}

\newlength{\figwidth}
\newcommand\ltappr{{{\lower4pt\hbox{$<$} } \atop \widetilde{ \ \ \ }}}

\newlength{\bxwidth}
\bxwidth=\columnwidth

\newcommand{\braket}[1]{\left\langle #1\right\rangle}
\newcommand{\pref}[1]{(\ref{#1})}
\newcommand{\abs}[1]{\left\vert #1 \right\vert}
\newcommand{\ua}{\uparrow}
\newcommand{\da}{\downarrow}
\newcommand{\be}{\begin{equation}}
\newcommand{\ben}{\begin{equation*}}
\newcommand{\ee}{\end{equation}}
\newcommand{\een}{\end{equation*}}
\newcommand{\bs}{\begin{split}}
\newcommand{\es}{\end{split}}
\newcommand{\bmx}{\begin{array}}
\newcommand{\emx}{\end{array}}
\newcommand{\bea}{\begin{eqnarray}}
\newcommand{\bean}{\begin{eqnarray*}}
\newcommand{\eea}{\end{eqnarray}}
\newcommand{\eean}{\end{eqnarray*}}
\newcommand{\dn}{^{\vphantom{\dagger}}}

\newcommand{\matl}[1]{\bmx{ll}#1\emx}
\newcommand{\matc}[2]{\left(\bmx{#1}#2\emx\right)}
\newcommand{\mat}[1]{\left(\bmx{cc}#1\emx\right)}

\usepackage{color}

\begin{document}

\title{Large-$N$ Approach to the Two-Channel Kondo Lattice}
\author{Ari Wugalter}
\affiliation{Center for Materials Theory, Rutgers University, Piscataway, New Jersey, 08854, USA} 
\author{Yashar Komijani}
\affiliation{Center for Materials Theory, Rutgers University,
Piscataway, New Jersey, 08854, USA} 
\author{Piers Coleman}
\affiliation{Center for Materials Theory, Rutgers University,
Piscataway, New Jersey, 08854, USA} 
\affiliation{Hubbard Theory Consortium, Department of Physics, Royal Holloway, University of London, Egham, Surrey TW20 0EX, UK}
\date{\today}

\begin{abstract}
This paper studies the two-channel Kondo lattice in
the large-$N$ limit at half-filling. In this model, 
the continuous channel-symmetry
is spontaneously broken, forming a channel-ferromagnet in which one
conduction channel forms a Kondo insulator, while the other remains
conducting. The paper discusses how 
this ground-state can be understood using the
concept of \emph{order fractionalization}, in which the channel magnetization breaks up into an emergent spinor order parameter.  By integrating out the
fermions we derive an effective action that describes this symmetry
breaking and its emergent collective modes. A remarkable observation is
that topological defects in the order parameter carry a $U(1)$ flux,
manifested in the Aharonov-Bohm phase picked by electrons that 
orbit the defect. By studying the effective
action,  we argue that the phase
diagram contains a non-magnetic transition between a large and a small
Fermi surface.
\end{abstract}
\maketitle

\section{Introduction}

The two-channel Kondo impurity and lattice models have a long history.  The
impurity version of this model was introduced by Blandin and
Nozi\`eres,\ \cite{Nozieres80}  who demonstrated that both the weak
and  strong-coupling fixed points of this model are unstable, 
flowing  to an intermediate coupling fixed point. 
This novel fixed point was later studied 
using 
Bethe ansatz,\,\cite{Natan,Tsvelik1984} conformal field
theory,\,\cite{Affleck92,Affleck}
bosonization,\,\cite{emery92,Sengupta94} numerical renormalization
group\,\cite{Pang91} and Majorana representation\,\cite{compactified}
establishing it as a  quantum-critical ground-state with non-Fermi
liquid properties and a fractional 
residual entropy $S=\frac{1}{2}k_{B}\ln 2$.

The lattice variant of this model, the two-channel Kondo lattice, was
proposed by Cox,\,\cite{Cox1996} as a quadrupolar Kondo 
description of the heavy fermion
superconductor UBe$_{13}$. Cox argued that the crystal-field-split $5f^{2}$ ground-state of UBe$_{13}$ is characterized by a non-Kramers
$\Gamma_{3}$ doublet, whose degeneracy is protected 
by cubic crystal symmetry, rather than the time-reversal symmetry of
Kramers doublets.  The key idea of Cox's model is that the criticality
of the single-impurity model will nucleate new forms of order in a lattice
environment.   Cox's 
two-channel Kondo lattice also forms the basis of proposed models for the
hidden order compound URu$_{2}$Si$_{2}$, where the entanglement
of a magnetic $\Gamma_{5}$ non-Kramers doublet with conduction electrons leads to the formation of a spinorial \emph{hastatic} order parameter.\,\cite{Chandra2013} 

The two channel Kondo lattice Hamiltonian 
\begin{eqnarray}\label{twoch}
H&=&\sum_{a=1}^2\sum\limits_{\vec{k}\alpha}\epsilon_{\vec{k}}c^\dagger_{\vec{k}\alpha
a}c_{\vec{k}\alpha a}\cr
&+& \sum\limits_{j\alpha\beta }\left(
J_{1}c^\dagger_{j\alpha 1}\vec{\sigma}_{\alpha\beta}c_{j\beta 1} 
+J_{2}c^\dagger_{j\alpha 2}\vec{\sigma}_{\alpha\beta}c_{j\beta 2} 
\right) 
\cdot\vec{S}_j,\label{eq2CKL}
\end{eqnarray}
defines the coupling between 
a lattice of local  moments 
$\vec S_j$ with two separate conduction seas, labelled by $a=1,2$, 
with coupling constants $J_{1}$ and $J_{2}$, respectively. 
The operator 
\begin{equation}\label{}
c\dg_{ja \alpha }=\frac{1}{\sqrt{{\cal N}_{s}}}\sum_{\vec k} e^{-i\vec k\cdot\vec
R_j}c\dg _{\vec{ k}a\alpha },
\end{equation}
creates an electron at site $j$, channel $a$, with 
spin component $\alpha $. 
Here ${\cal N}_{s}$ is the number of sites in the lattice. We are
particularly interested in the case of the 
symmetric two channel Kondo model, where $J_{1}=J_{2}$, which has 
channel exchange
symmetry $1\leftrightarrow 2$. Microscopically, this symmetry has its
origins in either time-reversal symmetry, or crystal point-group
symmetry. 
For example, in
a quadrupolar Kondo effect, the $\alpha $ are pseudo-spin orbital indices 
while the ``channel'' index is actually the spin
of scattered electrons, so that channel exchange symmetry is actually
time-reversal symmetry. 
In fact at $J_1=J_2$, the two channel Kondo lattice
develops an $SU (2)$ channel symmetry under which the Hamiltonian is
invariant w.r.t. continuous
rotations between the two channels. 

Two recent developments provide a  motivation to return to
this model. 
The recent discovery of
a new class of  ``1-2-20'' Praseodymium compounds,
with formula PrTr$_{2}$Al$_{20}$ (where Tr denotes a transition metal
ion Tr= Ti, V) or PrTr$_2$Zn$_{20}$ (where Tr=Ir, Rh) and a $4f^{2}$  ground-state appear to form a
new realization of Cox's original model.\,\cite{Onimaru2019,Worl19}
Unlike UBe$_{13}$, the smaller hybridization of
the Pr atoms makes it possible to definitively confirm the $\Gamma_{3}$
ground-state of these materials. Moreover, they exhibit a wide
variety of exotic ground-states, including triplet superconductivity, 
which appear consistent with novel patterns of entanglement between
the non-Kramers doublets and the conduction sea.

Our second motivation is conceptual.
Recent work\,\cite{Komijani18} has proposed an
interpretation of the expansion of the Fermi surface associated with
the Kondo effect as a manifestation of spin
fractionalization. This interpretation allows the Kondo effect to be
understood without attributing an anthropomorphic electronic 
origin to the neutral local moments, whose original origin as
microscopic qubits, whether electronic, nuclear or otherwise, is
entirely absent from the Kondo lattice Hamiltonian perspective. 
One of the interesting consequences of this
interpretation, is that it develops the phenomenon of 
``order fractionalization'', in which symmetry-broken ground-states acquire half-integer, spinorial character. 

A key property of the  two-channel impurity Kondo model, is that
its quantum critical ground-state is unstable to a variety of relevant, 
symmetry-breaking Weiss fields.\,\cite{Affleck92} In the lattice, this incipient quantum criticality gives way 
to a rich phase diagram of competing phases, providing an ideal
laboratory for studying the order fractionalization proposal. 
Dynamical mean-field theory calculations of the two-channel Kondo
lattice have reported an incoherent metal,\,\cite{Jarrell96}
odd-frequency pairing states as well as
antiferromagnetism.\,\cite{Jarrell1997} Recently, however, there are
various indications that the two-channel Kondo lattice also contains a
Fermi liquid phase in which the Kondo effect spontaneously develops in one
of the channels.\,\cite{Hoshino11,Hoshino13,Kuramoto,Zhang18} 
Experimental support for this phase is provided by the measurements on PrIr$_2$Zn$_{20}$.\,\cite{Onimaru2019} 
At high temperatures, this material displays non-Fermi liquid
properties, with temperature-dependent resistivity $\rho\sim \sqrt{T}$, expected from
dilute two-channel Kondo impurities. At lower temperatures, there is a
phase transition into a ``dome'' of Fermi liquid (FL),\,\cite{Onimaru2019} a
strong candidate for the channel symmetry-broken state.
At half-filling, this broken symmetry state results in a Kondo insulator in
one channel, leaving behind a conducting metal in the other. 
This state  is the main focus of the current paper.

\subsection{Spin fractionalization and Oshikawa's Theorem}\label{}

We begin by reviewing the key arguments for fractionalization in the
Kondo lattice, before going on to details of our current study. 
The prototypical single-channel Kondo lattice model is
\be
H=\sum\limits_{\vec{k}\alpha}\epsilon_{\vec{k}}c^\dagger_{\vec{k}\alpha}c_{\vec{k}\alpha}+J\sum\limits_{j\alpha\beta}
c^\dagger_{j\alpha}\vec{\sigma}_{\alpha\beta}c_{j\beta}\cdot\vec{S}_j.\label{eq1CKL}
\ee
where 
\begin{equation}\label{}
c\dg _{j\alpha }=\frac{1}{\sqrt{{\cal N}_{s}}}\sum_{\vec k} e^{-i\vec k\cdot\vec
R_j}c^\dagger_{\vec{ k}\alpha },
\end{equation}
creates an electron at site $j$. 
Although this model has a complex phase diagram, 
for sufficiently large Kondo coupling $J$ it realizes a
Fermi liquid (FL) in which the local moments are screened by
conduction electrons. 
%``Fermi liquid'' (FL). 
% In this phase,
% $M_j=\sum_{\alpha\beta}\langle{c^\dagger_{j\alpha}\vec{\sigma}_{\alpha\beta}c_{j\beta}\cdot\vec{S}_j}\rangle$
% acquires a finite value.
Numerical and analytical studies of
the model have shown that FL phase 
is distinct from the original conduction electron FL, for the Fermi surface
(FS) is enlarged, as if the local moments have delocalized as
electrons. 
This observation has
been placed on rigorous foundation by Oshikawa\,\cite{Oshikawa} who
argued, using a topological approach, that if the ground state
of \pref{eq1CKL} is a FL, 
it develops a large
FS, in which the volume of the Fermi surface $v_{FS}$ counts the
density of both
electrons {\it and} spins. 
\begin{equation}\label{eqOshi}
2 \frac{v_{FS}}{(2\pi)^{D}} = n_{e}+ n_{s},
\end{equation}
where $n_{e}$ and $n_{s}$ are respectively, the density of electrons
and local moments per unit cell.

\begin{figure}[tp!]  \includegraphics[width=\linewidth]{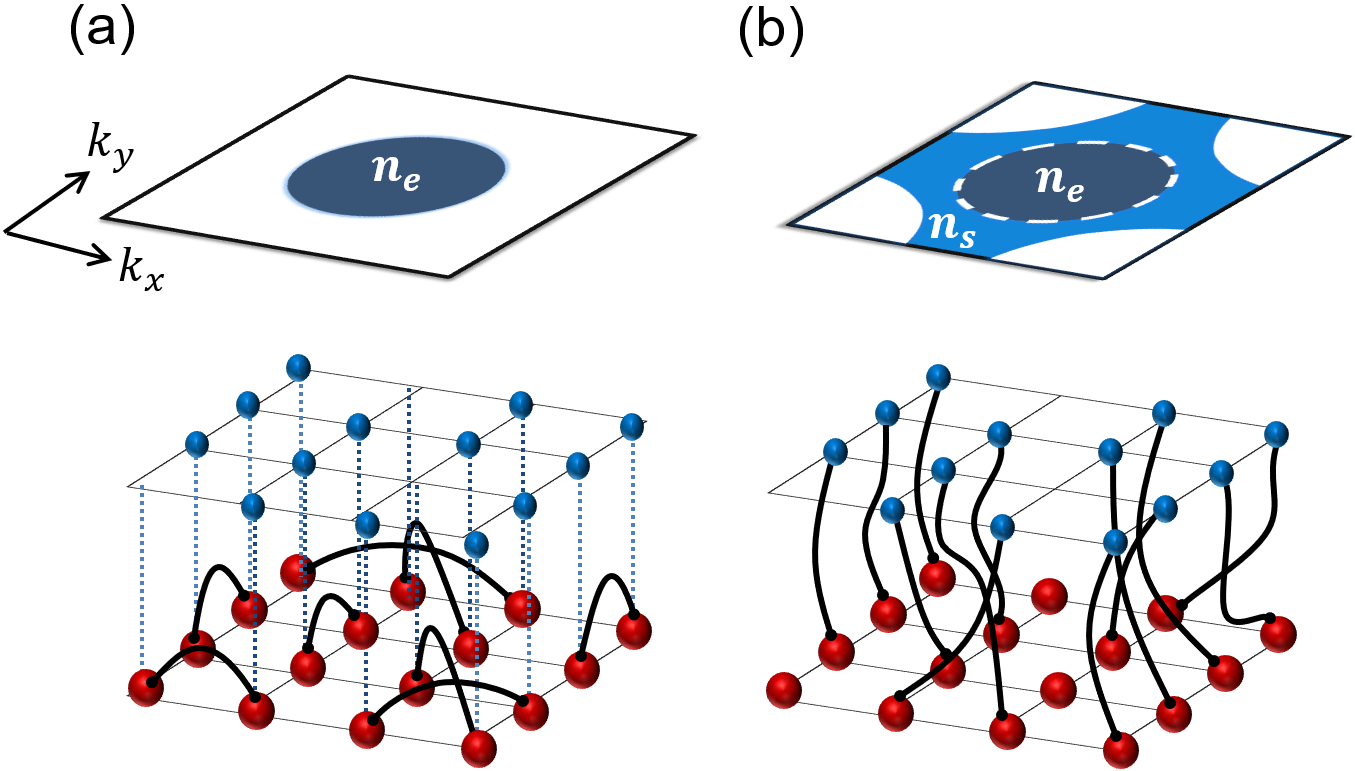}
\caption{\small\raggedright Schematic representation of a
single-channel metallic Kondo lattice (bottom) and corresponding Fermi
surface (FS) (top). Local moments and conduction electrons are
marked in red/blue. 
(a) At weak Kondo coupling, the local moments are magnetically correlated
and decoupled from the conduction sea, which has a small Fermi
surface.  (b) For large Kondo interaction, a paramagnetic state 
is formed in which the electrons and local moments bind into Kondo
singlets, forming a large Fermi surface. 
If there are less electrons than local moments, 
the unscreened moments form a gas of mobile holes.
At half-filling (not shown)
Kondo screening may drive a metal into an insulator.}\label{fig0}
\end{figure}

At half-filling, the expansion of the FS to fill the entire Brillouin
zone leads to a Kondo insulator.  
One of the ways to visualize this state is to consider
the strong coupling limit, where $J$ is much larger than the
bandwidth $W$ of the conduction electrons. 
When the number of spins and
conduction electrons are equal, a local singlet forms at each site,
with an insulating gap of size $J$. 
Hole doping away from half filling (Fig. \ref{fig0}) 
then gives rise to a small hole-like Fermi
surface of heavy electrons.  The volume of the FS counts both electrons and spins.

From a traditional stand-point, 
the FL phase and Kondo insulating phase 
of the Kondo lattice are the renormalized
counterparts of a FL and a band insulator, respectively. From
this traditional perspective, the Kondo lattice Hamiltonian
\pref{eq1CKL} is the result of a Wilsonian renormalization of an Anderson
lattice model, which describes hybridization between non-interaction
$c$-electrons and $f$-electrons with on-site Hubbard interaction $U_0$
\be H=\sum_k
\eps_kc\dg_{k\alpha}c\dn_{k\alpha}+\sum_j[V_0(c\dg_{j\alpha}f\dn_{j\alpha}+h.c.)+U_0n^f_{j\ua}n^f_{j\da}].\label{eqAnd}
\ee 
In the non-interacting limit the Anderson model has a large FS and as 
long as the interaction $U$ can be switched on adiabatically, forming a
Landau Fermi liquid, the FS volume will be unaffected.\,\cite{martin82}

However, the process of taking the low-energy limit of the Anderson model projects out empty and double-occupancy of $f$-electrons (corresponding to the lower and upper Hubbard $f$-bands),
reducing the four-dimensional Hilbert space of the physical 
electrons $f_j$ to the two-dimensional Hilbert space of the local moment
$\vec{ S}_{j}$.  The final Kondo model contains no trace of the
electronic origin of its local moments. 
Yet despite this irreversible loss of Hilbert space, 
{\it emergent} $f$-electron fields re-appear at
low-energies to expand the FS. Indeed, the high-energy origin of the local moments
is entirely irrelevant. The local moments could conceivably
even be nuclear in origin, antiferromagnetically coupled to electrons via a
hyperfine interaction, which if sufficiently large to overcome nuclear
magnetism, would also give rise to a large Fermi surface. 
This  extreme example makes it clear that the $f$-electrons which develop in the Kondo
lattice are emergent, 
independently of the spins' original microscopic origin. 

The large-$N$ mean-field theory using the Abrikosov fermion representation of the spin 
\be
S_{\alpha\beta} (j)\to f\dg_{j\alpha} \left(\frac{\vec{\sigma }}{2} \right)_{\alpha \beta }
 f\dn_{j\beta},\label{eqAbrikosov}
\ee
provides a simple interpretation of these
results,\,\cite{read83_1,Coleman} predicting that at low energies 
the product of local moment and conduction electron operators
behaves as an emergent $f$-electron field 
\begin{equation}\label{eqOPE}
J(\vec{\sigma}_{\alpha \beta }\cdot
\contraction{}
{\vec{S}}
{)}
{\psi }
\vec{S})
c_{\beta } =  V \hat f_{\alpha }.
\end{equation}
Here, the horizontal line contracting the spin and the fermion implies
that at long times, this 
combination acts as a single composite 
fermion. 

Eq.\,\pref{eqOPE}
can be regarded as an operator product identity in
the sense that the composite expression on the left can be replaced by
the expression on the right in long-time correlation
functions. 
The emergent amplitude $V$ and fermion $f$
are only defined modulo a $U(1)$ phase; an internal gauge degree of
freedom that implements the elimination of charge fluctuations of the
$f$-electrons. In
the Kondo ground state the internal gauge field locks to to the
external electromagnetic gauge field, providing the emergent $f$
electrons with an effective electromagnetic charge, which contributes
to the FS volume. 

Re-inserting Eq.\,\pref{eqAbrikosov} back into Hamiltonian \pref{eq1CKL},
we see that the formation of the fermionic bound-state implies that 
the low energy physics of a Kondo lattice is described by an Anderson model \pref{eqAnd}
with hybridization $V$ and an interaction $U\to 0$ that is zero in the large
$N$ limit.\,\cite{Hewson} However, if this behavior is independent of the high energy
origin of the Kondo physics (whether it describes an electronic or a
nuclear spin), we are obliged to interpret the equations \pref{eqAbrikosov}
and \pref{eqOPE} as a fractionalization of the
Kondo spin into emergent $f$-electrons. 
While the mean-field theory is only reliable in the
large-$N$ limit, recent numerical renormalization group (NRG) studies have shown that this interpretation applies to the Kondo impurity even for the case of spin-1/2 $SU(2)$ spins.\,\cite{Komijani18}

\subsection{Order fractionalization and Two-Channel Kondo Lattice}\label{}

The spin-fractionalization interpretation of the Kondo effect 
raises fascinating questions when applied to 
the two-channel Kondo lattice (Eq.\,\ref{twoch}). A formal application of Oshikawa's topological argument to this 
model simply leads to the conclusion that the total FS volume of the two channels is expanded by the spins, i.e
\begin{equation}\label{}
n_{e1}+n_{e_{2}}+n_{s}= \frac{2}{(2\pi)^{3}} (v_{FS}^{(1)}+v_{FS}^{(2)}).
\end{equation}
However, in order to form a Fermi liquid, the two channel Kondo
lattice needs to break the channel symmetry responsible for non-Fermi
liquid behavior. Blandin-Nozieres scaling arguments suggest
that if $J_{1}=J_{2}+\epsilon $, the asymmetry becomes relevant, and
the Kondo effect and the FS
expansion will develop exclusively in the strongest channel. 
In this channel asymmetric
state, 
\begin{eqnarray}\label{l}
n_{e1}+n_{s}&=& \frac{2}{(2\pi)^{3}} v_{FS}^{(1)},\cr
 \qquad n_{e2}&=& \frac{2}{(2\pi)^{3}} v_{FS}^{(2)}. 
\end{eqnarray}
and if $n_{e1}+n_{s}=2$, a Kondo insulator forms
exclusively in channel one.  
Since the second channel remains
conducting, we shall refer to this state as a ``half Kondo insulator''. Now suppose we restore the channel symmetry
by sequentially taking 
$\epsilon\to 0$ at each site in the lattice.  Those sites where 
the channel symmetry is restored will nevertheless 
feel a channel asymmetry derived from the channel polarization of the
Kondo singlets at neighboring sites. Like the Weiss field in a
magnet, this effect has the potential to 
preserve the channel magnetization in the ground-state, 
even when $\epsilon=0$ has been restored to zero at every site. 
Providing the Weiss fields
are channel-ferromagnetic, the ``half Kondo insulator'' will
survive the restoration of channel symmetry.  This then is an argument
for the development of a spontaneous broken channel symmetry. 

In this paper we examine this argument 
within the large $N$ expansion. Our results confirm the stability of
the ``channel ferromagnet'', a state with 
spontaneously broken channel symmetry and a 
``channel magnetization''
\begin{equation}\label{}
{\vec M} (x_{j}) = \braket{ c\dg _{ja\beta } {\vec\tau}_{aa'}\left(\vec{\sigma }_{\beta \delta }\cdot \vec{S}_{j} \right)c_{ja\delta}},
%\vec M(x_j)=\braket{s_{ja}\cdot \vec S_j}
\end{equation}
Here, $\vec\tau= (\tau_{1}, \tau_{2}, \tau_{3})$ are a set of Pauli
matrices in the channel space. $\vec M$ forms a vector in the channel Bloch sphere, indicating with which channel (or their linear combination) the spin forms the spin-singlet. 

However, the  channel symmetry breaking co-exists 
with the spin-fractionalization of the Kondo effect. In the case where 
$J_{1}>J_{2}$, the fractionalization of the spins involves the formation
of a bound-state in channel one, 
\begin{equation}\label{eqOPE1}
J(\vec{\sigma}_{\alpha \beta }\cdot
\contraction{}
{\vec{S}}
{)}
{\psi }
\vec{S})
c_{1\alpha} =  V \hat f_{\alpha }.
\end{equation}
However, for $J_{1}=J_{2}$ the presence of a perfect $SU (2)$ channel
symmetry implies that in the general channel-symmetry broken state, 
\begin{equation}\label{eqOPE2}
J(\vec{\sigma}_{\alpha \beta }\cdot
\contraction{}
{\vec{S}}
{)}
{\psi }
\vec{S})
c_{a\alpha} =  V z_{a} \hat f_{\alpha }.
\end{equation}
where $z_{a}$ is two component unit spinor. 
Hence, the hybridization of the two-channel Kondo lattice is now a
{\it two-component spinor} $V_{a}= Vz_{a}$.  The resulting channel magnetization 
$\vec M$ can then be represented
in terms of a 
fractionalized spinorial order parameter $z (x)$
\begin{eqnarray}\label{l}
\vec M (x) \propto %\biggl(
z\dg (x) \vec\tau z (x).%\biggr).
\end{eqnarray}

The development of an associated insulating behavior in one channel, 
implies that this is more than a simple
CP$^{1}$ representation of the channel magnetization. 
Conventional broken symmetries give rise to local, symmetry breaking
scattering potentials, such as the pairing field of an s-wave
superconductor, or the Weiss field of a ferromagnet
(Fig. \ref{weiss}a).  On 
length-scales larger than the order-parameter
coherence length $\xi$, the corresponding electron self energy is
local, e.g
\begin{equation}\label{}
\Sigma_{\alpha \beta } (1,2) = M_{\alpha \beta } (1)\delta_{\xi} (1-2)
\end{equation}
where $\delta_{\xi} (1-2)$ is a delta-function, coarse-grained on the
scale of $\xi$.  
While we can formally decompose $M_{\alpha \beta }\equiv \vec{M}\cdot \tau_{\alpha \beta }= V_{\alpha }\bar
V_{\beta }$ as a $CP^{1}$ product of spinors, in a conventional
ordering process the two spinors are confined, and 
always act together at a single
space-time point,  as a 
vectorial Weiss field. 

However, in the two channel Kondo lattice, the channel magnetization
does not create a local scattering potential. Instead, the electrons
scatter resonantly  off the screened local moments, a process
represented by the many-body hybridization with the $f$-fields 
that arise from the Kondo spin-fractionalization.   The electron 
self-energy that this gives rise to, is highly non-local in space-time, 
a self energy of the schematic form 
\begin{equation}\label{}
\Sigma_{\alpha \beta } (1,2) = V_{\alpha } (1)G_{f} (1-2)\bar V_{\beta } (2),
\end{equation}
where $G_{f} (1-2)$ is the bare $f$-electron propagator between $2$ and
$1$ (Fig. \ref{weiss}b). 
%For example, in
%momentum space, the channel asymmetric 
%electron self-energies take the form 
%\begin{equation}\label{}
%\Sigma_{\alpha \beta } (\bk ,\omega) =\frac{ V_{\alpha }\bar  V_{\beta
%}}{\omega- \epsilon_{f\bk }}
%\end{equation}
%where $\epsilon_{f\bk }$ is the dispersion of the unhybridized
%f-band. 
To form an insulator, the unhybridized $f$-band must lie within the
insulating gap, and the consequential absence of inelastic scattering
at these energies guarantees that the 
Green's function $G_{f} (1-2)$ is infinitely retarded in space and
time, so there is no coherence length-scale beyond which the  
two spinor variables $V (1)$ and $\bar V(2)$ coalesce into a
single vector order parameter. In this way, the channel magnetization
has fractionalized. 
\begin{figure}[htb]
{\includegraphics[width=\columnwidth]{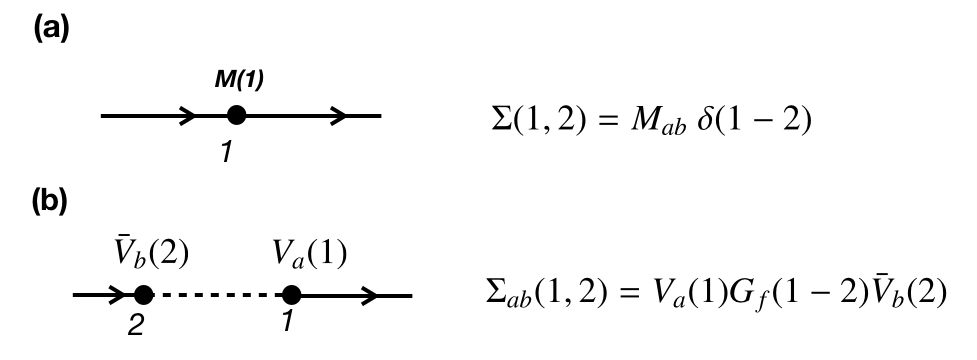}}
\caption{\raggedright Feynman diagrams for
(a)local scattering off a conventional Weiss order
parameter field $M (1)$ and (b) resonant scattering off a fractionalized
order parameter, where the dotted line represents the propagator of
the fractionalized spin.
}\label{weiss}
\end{figure}

\subsection{Collective modes and Topological Defects}

Another key interest in this paper is to examine
gapless modes of the two-channel Kondo lattice in the channel symmetry
broken phase, corresponding to the Goldstone rotations of the 
the three Euler angles of the spinor order parameter.  
We shall show that as in the single-channel Kondo lattice, 
one of these modes is absorbed by a Higgs mechanism that
locks the $U(1)$ gauge fields associated with the fractionalized $f$-electron
to the external electromagnetic field, giving the $f$-electrons physical
charge, and driving the 
large electronic FS. 

A key finding, is that 
the channel magnetization $\vec M$ admits topological configurations: 
skyrmions in
2 dimensions or `hedgehog' instantons in 2+1 dimensions, which couple
to the underlying 
gauge charges in the system. These topological excitations modify the
electronic spectrum.
When the Kondo temperature becomes sufficiently weak, 
the proliferation of such hedgehog defects is expected to lead 
to 
a `quantum disordered' phase, in which the coherence between the
gauge field of spinons and the external field is destroyed and the ground state has a small FS.

\begin{comment}
In the mean-field theory, this the channel magnetization 
factorizes in terms of a channel spinor $V_{a}$
\begin{equation}\label{}
{\vec n} ({x_{j}})\sim V\dg_j\vec{\tau }V_{j}
\end{equation}
where $V_{a}= (V_{1},V_{2})^{T}$ determines the linear combination
of channels which selectively forms a singlet (``hybridizes'')
with the local moments. 
This product structure 
suggests that the composite order fractionalizes. 
A separate indication of this fractionalization appears in the
factorized structure of the mean-field
electron self-energy
\begin{equation}\label{}
\Sigma_{aa'} (\tau_{2},\vec{R}_{i};\tau_{1}, \vec{R}_{j})
=\left(\frac{\delta_{ij}}{2}{\rm sgn} (\tau_{1}-\tau_{2}) \right)
V_{a} (2) V^{*} _{a'} (1).
\end{equation}
\end{comment}

The structure of the paper is as follows. In section \ref{meanfield}
we use the large-$N$ mean-field theory to study the ground state of
the two-channel Kondo lattice. In subsection \ref{ofc} we show that a
natural description of the ground state in the ordered phase is
provided by the concept of order
fractionalization.\,\cite{Komijani18} In section \ref{collective}, we integrate out the fermions and   derive an effective action that
describes collective excitations of the system, including the Higgs
mechanism and the small to large FS transition mentioned above. Finally,
we conclude the paper in section \ref{conclusion} and list a number of
open questions. A number of appendices are included to provide
additional derivations and details used in the paper.

\section{Mean Field Theory of the two-channel Kondo Lattice}\label{meanfield}
We consider a two-channel Kondo lattice, represented by Hamiltonian
\pref{eq2CKL}. 
As 
written, the channel index is an orbital quantum number, while the
local degrees of freedom are spins. We note that in the equivalent
quadrupolar 
formulation of the two-channel Kondo model proposed by Cox, $\vec{
S}_{j}$
represents a non-Kramers doublet. In this case, $\alpha $ is a
quadrupole index while the channel index corresponds to the
``up'' and ``down'' spins of the conduction sea.
To develop a controlled mean field theory, we extend the number of spin components from $2$ to $N$ by taking the spins from an irreducible representation of $SU(N)$ instead of $SU(2)$.
This  generalized
version of the model uses the Coqblin-Schrieffer form of the interaction
\begin{equation}
{H}=\sum\limits_{\vec{k}a\alpha}\epsilon_{\vec{k}}c^\dagger_{\vec{k}a\alpha}c_{\vec{k}a\alpha}+J\sum\limits_{ja\alpha\beta}c^\dagger_{ja\alpha}c_{ja\beta}S_{\beta \alpha } (j).
\end{equation}  
Here, $S_{\beta \alpha } (j)$ are representations of generators of the $SU (N)$ group.
This symmetric two channel Kondo lattice model possesses an
$SU_{spin}(N)\times SU_{channel}(2)\times U_{charge}(1)$ symmetry.
We shall use an Abrikosov fermion representation of the $SU(N)$
spin operators, 
\begin{eqnarray}\label{l}
S_{\alpha \beta } = 
f\dg_{\alpha }f_{\beta } - \frac{Q}{N}\delta_{\alpha \beta },
\end{eqnarray}
subject to the constraint
\begin{equation}
\label{}
n_{f} = f\dg_{\alpha }f_{\alpha } = Q,
\end{equation}
\\
where  $Q$ is an integer. % which sets the number of elementary spins that make up the higher antisymmetric representation. 
To develop a controlled large-$N$ expansion for the Kondo lattice, the coupling constant is rescaled by a factor of $1/N$ to guarantee that each term in the Hamiltonian scales extensively with $N$. In terms of
the Abrikosov representation, the Hamiltonian becomes
\bea
{H} &=&
\sum\limits_{\vec{k}a\alpha}\epsilon_{\vec{k}}c^\dagger_{\vec{k}a\alpha}c_{\vec{k}a\alpha} 
-\frac{J}{N}\sum\limits_{ja\alpha\beta}(c^\dagger_{ja\alpha}f_{j\alpha})(f^\dagger_{j\beta}c_{ja\beta})
\\
&&+\sum_j\lambda_j(n_{fj}-Q)
\eea
where the Lagrange multiplier $\lambda_{j}$ is introduced to impose the
constraint $n_{f}=Q$ at each site. The Abrikosov factorization of the
spin operator permits one to write the partition function as a path
integral
\begin{equation}\label{}
Z = {\rm Tr}\left[e^{- \beta  H} \right]= \int {\cal D}[\bar  c,
c,\bar f, f, \lambda]e^{-S}.
\end{equation}
Inside the path integral 
the  interaction can be decoupled in each channel 
using a Hubbard-Stratonovich transformation, 
\bea\label{l}
-\frac{J}{N}&&\sum _{\alpha\beta}(c^\dagger_{ja\alpha}f_{j\alpha})(f^\dagger_{j\beta}c_{ja\beta})\cr
&&\qquad\qquad\rightarrow
\Big[(c\dg _{ja\alpha} f_{j\alpha}){V}_{a j}+ {\rm H.c}\Big]
+\frac{N}{J}|V_{a j}|^2,\qquad\nonumber
%&=&{c}\dg _{ja\alpha} ({V}_{a j}f_{j\alpha})+{\rm H.c}+\frac{N}{J}|V_{a j}|^2,
\eea
where the 
``hybridization'' field $V_a$ is to be integrated over inside the
path integral, 
\begin{displaymath}
Z = \int\mathcal{D}[\overline{c},c,\overline{f},f,\overline{V}_a,V_a,\lambda]\ e^{-S}
\end{displaymath}
with the action
\begin{widetext}
\be
S=\int\limits^\beta_0
\mathrm{d}\tau
\Biggl\{
\sum\limits_{\vec{k}a}\overline{c}_{\vec{k}a}(\partial_\tau+\epsilon_{\vec{k}})c_{\vec{k}a}
+\sum\limits_{j\alpha}\overline{f}_{j}(\partial_\tau+\lambda_j)f_{j}
+\sum\limits_{ja}\biggl[
\overline{c}_{ja} (V_{a j}f_{j})+ (\bar f_{j} \bar  V_{aj})c_{ja}\biggr]+\sum_{ja }
\frac{N}{J}|V_{a j}|^2
-\sum_j \lambda_j Q
\Biggr\}.\label{eqsameJ}
\ee
\end{widetext}
Here, and in the following the summation over the spin indices $\alpha=1\dots N$ is implicit. 
\subsection{Symmetries and Gauge Transformations of the Hamiltonian}
The symmetric two-channel Kondo lattice exhibits a 
number of global and local symmetries: 
The conduction electrons are invariant under global { $U (1)\times SU (2)$} rotations in channel space\begin{displaymath}
c_{ja}\rightarrow g_{aa'}c_{ja'},\qquad
V_{a j} \rightarrow g_{aa'}V_{a 'j}.
\end{displaymath}
The $f$-electrons possess a local $U(1)$ gauge invariance associated with
the conserved $f$-charge $n_{f} (j)=Q_{j}$,
\begin{eqnarray}\label{}
f_{j}&\rightarrow& e^{i\chi_j}f_{j},
\quad V_{a j}\rightarrow  V_{a j}\ e^{-i\chi_{j}},\quad
\lambda_j\rightarrow \lambda_{j} (\tau ) - i \partial_{\tau }{\chi}_j (\tau ).\nonumber
\end{eqnarray}
In the single-channel Kondo impurity/lattice, this gauge
transformation is used to make the hybridization real with the price
of transforming the original static $\lambda_{j}$ into a dynamical,
time-dependent field.\,\cite{Coleman} 

It is convenient to represent the hybridization fields as a spinor
\begin{equation}
\pmat{V_{1} (j,\tau )\cr V_{2} (j,\tau )}\equiv  V (j,\tau ) \pmat{z_{1} (j,\tau ) \cr z_{2} (j,\tau ) }
\end{equation}
where $V(j,\tau )$ is a positive real number
representing the magnitude of the hybridization and 
$z_{1}$ and $z_{2}$ define a unit spinor with $z\dg z =
|z_{1}|^{2}+|z_{2}|^{2}=1$ which can be written in terms of Euler angles
\begin{equation}
z\equiv\pmat{z_{1} \cr z_{2} }
= 
e^{i \varphi/2}\pmat{
 \cos\theta/2\\
-e^{i\phi}\sin\theta/2 }.
\end{equation}
Since the underlying channel symmetry is $SU (2)$,  
the full range of values in this unit spinor involve a double
covering of the $SO (3)$ group, incorporated by doubling the range  
of $\varphi \in [0,4 \pi]$.

\begin{comment}
The hybridization field in the action plays the role of a Higgs 
field. The Anderson-Higgs effect 
allows us to use the local $U (1)$ gauge invariance to fix the gauge, 
absorbing the $U (1)$ phase $\varphi_{j} (\tau )$
into the constraint field, 
$\lambda_{j} (\tau)= \lambda_{j}+ i \partial_{\tau  }\varphi_{j}$, 
so that the action then takes the form, 
\begin{align}
S=\int\limits^\beta_0
\mathrm{d}\tau\Biggl\{\sum\limits_{\vec{k}a\alpha}\overline{c}_{\vec{k}a\alpha}(\partial_\tau+\epsilon_{\vec{k}})c_{\vec{k}a\alpha}+\sum\limits_{j\alpha}\overline{f}_{j\alpha}(\partial_\tau+\lambda_j)f_{j\alpha}+\sum\limits_{j\alpha}V_j\left[
(\overline{c}_{j1\alpha}u_{j}+
\overline{c}_{j2\alpha}e^{i\alpha_j}
v_{j})f_{j\alpha}+{\rm H.c.}\right]+\nonumber\\
+\sum_i\frac{NV_j^2}{J}-\sum_j \lambda_j Q \Biggr\}.
\label{eq:action}
\end{align}
The presence of the constraint field breaks the $U (1)$ symmetry, and
the resulting Higgs' effect gives rise to a phase stiffness in the
field $\phi_{j} (\tau )$ and an associated mass to the gauge invariant
constraint field $\lambda_{j}+i\partial_{\tau }\phi_{j}$.
\end{comment}

\subsection{Uniform Mean Field Solution}

\begin{figure}[tp!]
\includegraphics[width=\linewidth]{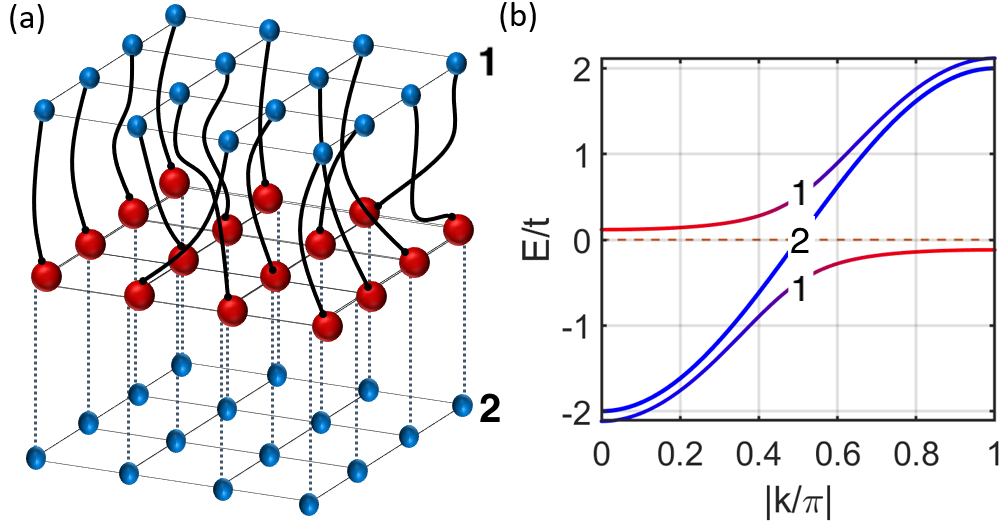}
\caption{\small\raggedright (a) Channel ferromagnetic state: the localized spins (red) hybridize with one out of two conduction channels (blue). (b) A schematic representation of the band structure in a half Kondo insulator.}\label{fig:FM}
\end{figure}

In the limit $N \rightarrow \infty$ the path integral is dominated by
the stationary points of the action characterized by
static, uniform configurations of the hybridization spinor, such that 
$V_{{j}} = V$, $\theta_{j}=\theta$, $\phi_{j}=\phi$,  $\varphi_{j}=\varphi$
and $\lambda_{j}=\lambda$ are all constant. 
The overall phase
$e^{i\varphi/2}$ in the hybridization
can be absorbed by a gauge transformation $f_{j}\rightarrow
e^{i\varphi/2}f_{j}$
of the $f$-electrons.
Moreover, by rotating in channel space
\begin{gather}
\begin{pmatrix}c'_{k1} \\
c'_{k2} \\
\end{pmatrix}=\begin{pmatrix}
\cos\theta/2
 & e^{-i\phi} \sin\theta/2\\
-e^{i\phi}\sin\theta/2& \cos\theta/2 \\
\end{pmatrix}\begin{pmatrix}c_{k1} \\
c_{k2} 
\end{pmatrix},
\label{eq:gfix}
\end{gather} 
the mean-field action becomes 
\begin{widetext}
\bea
S=\int\limits^\beta_0
\mathrm{d}\tau
\sum\limits_{\vec{k}\alpha}
\Biggl\{
( \bar c'_{\vec{k} 1}, \bar  f_{\vec{k} })
\left [ 
\partial_{\tau } +
\pmat{\epsilon_{\vec{k}}&V\cr V & \lambda}
\right]
\pmat{c'_{\vec{k}1 }\cr f_{\vec{k} }}
+
\overline{c}'_{\vec{k}2}(\partial_\tau+\epsilon_{\vec{k}})c'_{\vec{k}2
 }
+{\cal N}_{s}\left( \frac{NV^{2}}{J}-\lambda Q\right)
\Biggr\},\qquad
\label{eq:action}
\eea
\end{widetext}
where we have defined $f_{\vec{k} }= \frac{1}{\sqrt{\cal
N}_{s}}\sum_{j}f_{j }e^{i \vec{k}\cdot \vec{ R}_{j}}$. For the case
where $Q=N/2$ and a particle-hole symmetric conduction band, we have $\lambda=0$ by the symmetry. 

In this basis, the second conduction electron channel
decouples from the $f$-electrons as shown schematically in
Fig.\,\ref{fig:FM}(a).  
The
system reduces to a ``half Kondo insulator'', with 
a first hybridized channel, forming a fully gapped Kondo insulator with upper and lower bands dispersing according to
\be
E^{\pm}_{\vec{k}} 
=\frac{\epsilon_{\vec{k}}}{2}
\pm
\sqrt{ \left(\frac{\epsilon_{\vec{k}}}{2} \right)^2+V^2},
\ee 
and a second decoupled 
conduction band with dispersion $\epsilon_{\vec{k}}$. This is shown in Fig.\,\ref{fig:FM}(b). In two spatial dimensions, the free energy per site per particle is

\bea
\frac{F}{{\cal N}_s N}=
-T
\int\frac{\mathrm{d}^2k}{(2\pi)^2}
&\Big\{& 
\log\left[1+e^{-\beta {\epsilon_{\vec
{k}}}}\right]\nonumber\\
&&\hspace{-0.5cm}+
\sum_{\pm }
\ln
\left[
1+e^{-\beta E^{\pm }_{\vec{k}}}
\right]
\Big\} +\frac{V^2}{J}.\qquad\quad
\eea

\begin{figure}[t!]
    \centering    
    \includegraphics[width=\linewidth]{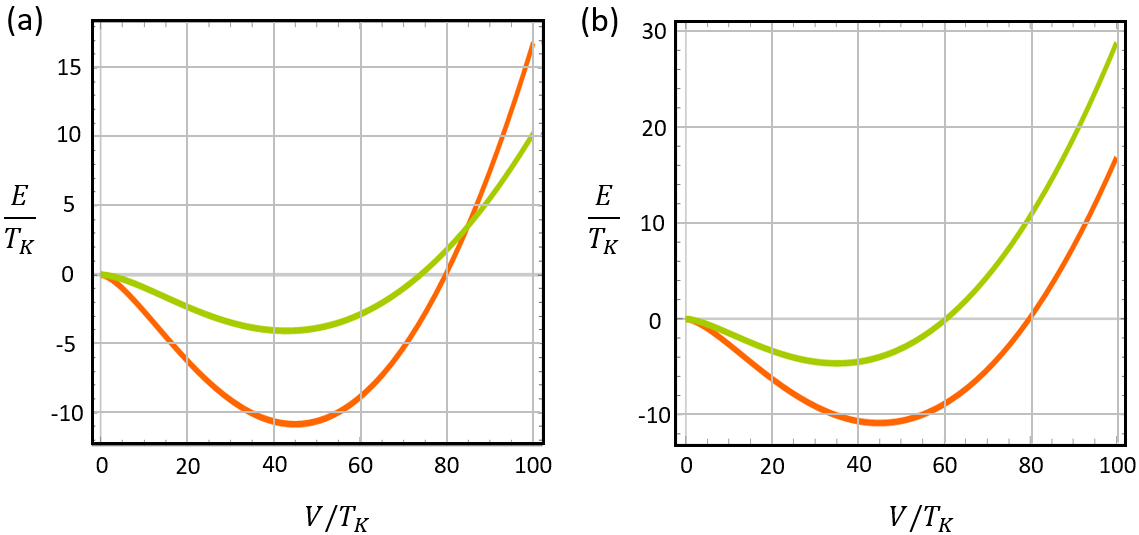}
    \caption{\small\raggedright (a) Ground state energy of a two-dimensional half-Kondo insulator state using constant (green) and exact (red) density of states. (b) Comparison of ground state energies of the half-Kondo insulator (channel ferromagnet in red) and the staggered hybridization solution (channel antiferromagnet in green). }
    \label{fig:energy}
\end{figure}

 The solution for the magnitude of the hybridization $V$ can be obtained self consistently from the stationarity condition
\begin{gather}
\frac{1}{{\cal N}_sN}\frac{\delta F}{\delta V^2}
=\frac{1}{J}-
\int\frac{\mathrm{d}^2k}{(2\pi)^2}\left(\frac{
f (E^{-}_{\vec{ k}})
-f (E^{+}_{\vec{ k}})
}{
2\sqrt{\left(\frac{\epsilon_{\vec {k}}}{2}\right)^2+V^2}}
 \right)
=0,\nonumber
\end{gather}
where $f (E)= [1+ e^{\beta E}]^{-1}$ is the Fermi-Dirac function. 
Using a constant density of states $\rho(\eps)
=\rho \theta(4t-\abs{\eps})$, where $\rho =1/8t$, we can solve for the hybridization
at $T=0$, giving
\begin{equation}\label{explicit}
\frac{1}{J}= \rho
\int_{-4t}^{4t}\frac{d\epsilon}{\sqrt{\epsilon^{2}+4V^{2}}}= 2\rho \sinh^{-1}\left(\frac{2t}{V} \right)
\end{equation}
or
\begin{equation}
V^{-1}=4\rho\sinh\left(\frac{1}{2\rho J}\right),
\end{equation}
or $V= 4t e^{-\frac{1}{2\rho J}} $ in the large band-width
limit. 
The ground state energy is plotted for different hybridization strengths for a constant density of states and for the exact density of states for a conduction band with nearest neighbor hopping in figure \ref{fig:energy}(a). The minimum of the ground state energy corresponds to the actual hybridization strength.

Similarly, the mean-field transition temperature $T=T_{K}$ is determined by 
\begin{equation}\label{}
\frac{1}{J\rho}= \int_{-4t}^{4t} d\epsilon \frac{1/2-f
(\epsilon)}{\epsilon} = \ln  \left[\frac{4t}{T_{K}}\left(
\frac{e^{-\psi (1/2)} }{4\pi}
\right) \right]
\end{equation}
From which it follows that 
\begin{equation}
T_K = 2t
\overbrace {\left(\frac{e^{-\psi (1/2)} }{2\pi} \right)}^{1.13}e^{-\frac{1}{\rho
J}} = \left(\frac{1}{3.53} \right) \left( \frac{V^{2}}{2t}\right)
\end{equation}
Note that while the direct gap is determined by $V$, 
the mean-field transition temperature, or Kondo temperature
is determined by the much smaller, indirect gap,
$\Delta_{g}=V^{2}/2t=3.53 T_{K}$.
In Appendix \ref{afm} we compare this half Kondo insulator state with
a channel symmetry breaking pattern, for which the hybridization is
staggered in channel space. The ground state energies of the half
Kondo insulator state and the staggered hybridization state are
compared in figure \ref{fig:energy}(b). We see that 
the half-Kondo insulator state is energetically stable with respect to
the channel antiferromagnet for all values of the coupling. This
is an important result, for it confirms that the channel Weiss fields
acting between different spin sites are ferromagnetic in nature at
half filling.

\subsection{Kondo-Heisenberg system}\label{KHS}

So far, we have assumed that the spins do not directly interact with
each other and each spin is a self-conjugate $Q=N/2$ fermionic
representation of $SU(N),$ so that together with a half-filled
conduction band the whole system has particle-hole
symmetry. More generally, the Kondo
coupling induces an RKKY interaction amongst the spins, which is
generically long-ranged and varies in space. Assuming that
the system does not magnetically order at low temperatures, in the
rest of the paper, we generalize our model $H\to H+H_{\rm AFM}$ to include
effects of a frustrated antiferromagnetic interaction among the spins:
\be
H_{\rm AFM}=\sum_{i,j}J^H_{ij}\vec S_i\cdot\vec S_j.
\ee
In the SU(N) limit, 
using the Abrikosov fermion representation, and applying the constraint this leads to
\be
H_{\rm AFM}=-\frac{1}{N}\sum_{(i,j), \alpha, \beta }J^H_{ij}\Big(f\dg_{i\alpha}f\dn_{j\alpha}\Big)\Big( f\dg_{j\beta}f\dn_{i\beta}\Big)
\ee
which can be decoupled by a Hubbard-Stratonovitch transformation $t_{ij}=\abs{t_{ij}}e^{i\varphi_{ij}}$
\bea
H_{\rm AFM}\to&&\sum_{(i,j)}\Big\{\frac{N\abs{t_{ij}}^2}{J^{H\dn}_{ij}}+
\Big[(\abs{t_{ij}}e^{i\varphi_{ij}}f\dg_{i\alpha}f\dn_{j\alpha}+h.c.)\Big]\Big\}.\qquad
\eea
with an implied summation over the repeated spin variables $\alpha \in[1,N]$.
We assume that the underlying spin fluid is a $U(1)$ spin liquid, characterized by a spinon Fermi surface. The phase factor
\be
e^{i\varphi_{ij}}\sim \exp\Big[{i\int^{\vec{R}_i}_{\vec{R}_j}{\vec A^{f}\cdot\vec dl}}\Big],
\ee
is in fact the Peierls substitution of a slowly varying $U(1)$ gauge degree of freedom. While the phases $\varphi_{ij}$ themselves are gauge-dependent, e.g. by a redefinition of $f_j$ that leaves the spin representation unaffected,
\be
f_j\to f_j e^{i\chi_j}, 
\qquad \varphi_{ij}\longrightarrow\varphi_{ij}-\chi _i+\chi_j,
\ee
the sum of the 
phases around plaquettes is gauge-invariant and corresponds to the $U(1)$ gauge flux 
through the plaquettes:\,\cite{Marston89}
\begin{equation}\label{}
\sum_{( i,j)\in\square}\varphi_{ij}\equiv \oint_{\square}
\vec{A}^{f}\cdot d\vec{l}= \Phi_{f}
\end{equation}
The combination of the magnitude/phase of $t_{ij}$ gives rise to a dispersion for the spinons $\eps^f_k$ so that the momentum-space Hamiltonian is  
\be
H_{\rm AFM}\to \sum_{ij}\frac{\abs{t_{ij}}^2}{J^H_{ij}}+\sum_{k\alpha} \eps^f_kf\dg_{k\alpha}f\dn_{k\alpha}.\label{eqKondoHeisenberg}
\ee
The simplest case is a nearest-neighbor tight-binding lattice of $q=1/2$ moments and no flux per plaquette $\sum_{( i,j)\in\square}\varphi_{ij}=0$, with the 2D spinon dispersion
\be
\eps_k^f=-2t_f(\cos k_x+\cos k_y).\label{eqspinondispersion}
\ee
If we allow the gauge fields $\vec{A}^{f}$ to vary slowly in space,
then the coarse-grained action of the $f$-electrons takes the form\,\cite{Marston89}
\begin{equation}\label{}
S_{f} = \int d^{3}xdt \bar f\left[-i \partial_{t} +\lambda +
\epsilon^{f} ({-i\vec{\nabla }+\vec{ A}})\right]f
\end{equation}
where we have replaced $\partial_{\tau }\rightarrow -i \partial_{t}$. 
This action is invariant under the gauge transformation $f\rightarrow
e^{i\chi }f$ and 
\be\label{l}
\lambda\rightarrow \lambda+ \partial_{t}\chi, \qquad
\vec{ A}^{f}\rightarrow  \vec{ A}^{f}+ \vec{\nabla }\chi,
\ee
allowing us to combine $A^{f}_{\mu}\equiv  (\lambda, \vec{A}^{f})$
into a single $U (1)$ gauge field that transforms as
$A^{f}_{\mu}\rightarrow A^{f}_{\mu}+ \partial_{\mu}\chi $. Hence, variations of $U(1)$ gauge
on top of the mean-field background can be taken into account by the 
by the minimal coupling $-i\partial_{\mu}\to  -
i\partial_{\mu}+A^{f}_{\mu}$ or  $p_\mu\to p_\mu+A^f_\mu$.

\subsection{Order fractionalization}\label{ofc}
In the two channel case, the $f$-electrons can be integrated out and the self-energy for the conduction electrons has the form
\be
\Sigma_{a b} (\vec{k}, \omega) =\frac{V_{a}\bar  V _{b}}{\omega-\eps^f_k},\label{eqselfenergy}
\ee
where $a,b=1,2$ are the channel indices of the conduction
electrons. 
Writing
$V_{a}= V z_{a}$, where $z\dg  z=1$ is a unit spinor, we have
\bea\label{l} \Sigma_{a b} (\vec{k}, \omega) =\frac{V^{2}}{\omega-\eps^f_k}
z_{a}\bar  z_{b}
=\frac{V^{2}}{\omega-\eps^f_k} \left( \frac{1 + \vec{
n}\cdot\vec{\tau} }{2} \right)_{ab} 
\eea where $\hat n = z\dg
\vec{ \tau }z$. 
Were it not for the strong frequency dependence of this 
self-energy, we could simply regard this term as a Weiss scattering
field created by a channel magnetization.

For a slowly varying order parameter, the self-energy becomes
\be\label{}
\Sigma_{a b} (2;1)
=
V_{a} (2) G_f(2-1)\bar V _{b} (1).
\ee
where $G_f(2-1)$ is the bare propagator of an $f$-electron 
from 1 to 2. 
To make the state insulating, the unhybridized f-band must 
cut the Fermi energy  to repel the conduction band from the
Fermi energy. This causes  $G_{f} (2-1)$ to develop infinite range
correlations in time, so that we are forced to regard the spinors $V_{a} (2)$ and $\bar V_{a'} (1)$ as independent variables. 

Part of this propagator is the
dynamic phase accumulated from 1 to 2. For example,  for non-dispersing $f$-electrons,
\be
G_f(2-1)\sim\exp \left[ -i\int_{t_{1}}^{t_{2}} \lambda (t')dt'\right].
\ee
At particle-hole symmetry $\lambda=0$, the self-energy in real space/time becomes
\be 
%\lim_{\abs{\tau_1-\tau_2}\to\infty}
\Sigma_{ab}(2;1)=-
\frac{\delta_{\vec{x}_{2},\vec{x}_{1}}
}{2}V_a (\tau_{2})
{\rm sgn}(\tau_2-\tau_1)
\bar V_b (\tau_{1})
.  \ee
More generally however, the $f$-state will develop a dispersion due to the
magnetic interaction
\begin{equation}\label{}
G_{f}(\vec{k},\omega) = \frac{1}{\omega - \epsilon^{f}_{k}}.
\end{equation}
In general, $\epsilon_{\vec{k}}^{f}$ has zeros and will cut
the Fermi energy on a surface $\{S_{0}: \vec{k}=\vec{k}_{0} \}$. 
Since the self-energy diverges at $\omega=0$ on this surface, 
it follows that $S_{0}$ corresponds to the zeros of the conduction electron propagator.
The corresponding
real-time propagator will take the form 
\begin{equation}\label{}
G_{f} (\vec{x},t)= \frac{e^{i{ k}_{0}x}}{x-v^{f}_{\vec{ k}_{0}}t}
\end{equation}
where $\vec{k}_{0}$ is at the point on the null surface $S_{0}$
with normal parallel to  the separation vector $\vec{ x}= x
\hat n$, so that 
\begin{equation}\label{}
\Sigma_{ab} (2; 1) = V_{a } (2)\frac{e^{i {
k}_{0}|\vec{x}_{2}-\vec{x}_{1}|}}{|\vec{x}_{2}-\vec{x}_{1}|
-v^{f}_{\vec{ k}_{0}} (t_{2}-t_{1})}\bar  V_{b } (1)
\end{equation}
In  conventional broken symmetry phases, the 
self-energy is local on a scale of the coherence length: but here, 
the resonant scattering process through an intermediate
spin-fluid means that the initial and final 
hybridization events 
$V_{\beta } (1)$ and $\bar  V_{\beta } (2)$, 
can be arbitrarily
separated in space and time. This is a key signature of the fractionalization. 
Notice that while the fermion field  $f$ 
and the hybridization order parameters  appearing here, are
only defined modulo a gauge transformation, the self-energy $\Sigma
(2-1)$  is invariant under these transformations. 

\section{Collective excitations}\label{collective}

\subsection{The soft modes}
\label{softmodes}

Within the large-$N$ mean-field theory, 
the Kondo coupling reduces to a hybridization between Abrikosov fermions and the two conduction bands, described by an spinor $V_a(x,\tau)$ in the Hamiltonian density
\be
{\cal H}_{int}\to \frac{V^2}{J_K}+(c\dg_a V_a f+h.c.).\label{eqHint}
\ee
At high temperature, the hybridization is strongly fluctuating, 
but once $T\ltappr T_K$, the hybridization spinor 
acquires a non-zero expectation value  $V_{a}\neq 0$. Longitudinal
fluctuations in the magnitude of $V_{a}$
are massive and gapped, but (transverse) fluctuations in the direction of the 
$V_a$ spinor develop soft modes. This physics 
is conveniently shown by writing $V_a(x,\tau)=Vz_a(x,\tau)$ and 
\begin{equation}\label{eqHyb}
 z(x,\tau)=\bb g(x,\tau)\mat{1\\0},
\end{equation}
where $ \bb g\in U(2)/U(1)\sim SU(2)$. We can parameterize $ \bb g$ by the three Euler angles
\bea
{ \bb g}(x,\tau)&=&e^{i\phi\tau_3/2}e^{i\theta\tau_2/2}e^{i\varphi\tau_3/2}.\label{eqgparam}
\eea
By integrating out the fermions, we can derive a long-wavelength effective action that describes the spontaneous symmetry breaking from this fluctuating phase to the channel ferromagnetic ground state.

In the following, we consider a long-wavelength ($k\sim 0$) approximation to the model \pref{eqKondoHeisenberg} and assume that the spinon dispersion is quadratic near $k\sim 0$. Moreover, we assume a continuum theory of conduction electrons with parabolic dispersions. This combination describes the low-energy limit of the Heisenberg-Kondo two-channel Kondo lattice, in presence of both, an external electromagnetic vector potential $A^{ex}$ and an internal gauge potential $A^f$. That such a continuum Kondo insulator exists, is discussed in Appendix \ref{continuum}.

The Hamiltonian density is ${\cal H} (x) ={\cal H}_0 (x) +{\cal H}_{int} (x)$ where ${\cal H}_{int}$ is given in eq.\,\pref{eqHint} and ${\cal H}_0$ is given by
\bea
{\cal H}_0 (x)&=&C (x)\dg\Big[\sum_{\nu=1}^d\frac{(p_\nu-eA_\nu^{ex})^2}{2m_c}+ieA_\tau^{ex}-\mu\Big] C (x)\nonumber\\
&+&f\dg (x)\Big[-\sum_{\nu=1}^d\frac{(p_\nu+ A^f_{\nu})^2}{2m_f}+iA^f_\tau+\lambda \Big]f (x).\qquad
\eea
The Wick rotation of scalar potentials is $A_\tau\to -iA_t$. Note that
the effective masses of the c- and f- bands are opposite to one-another,  to insure that
the hybridized channel is insulating. 
Here, $p_\nu=-i\partial_\nu$, $C=(\matl{c_1, & c_2})^T$ and the 
spin indices $\alpha\in[1, N]$ have been suppressed. The second line
describes a gapless $U(1)$ spin-liquid. The internal gauge field $A^f$
arises from decoupling the Heisenberg magnetic interaction between
$f$-electrons (spinons) as described in section (\ref{KHS})\cite{Marston89}. The temporal and spatial variations of the hybridization in Eq.\,\pref{eqHyb} can be absorbed into the $C$ electrons by a $C\to  \bb gC$ transformation. This leads to
\be
C\dg\partial_\mu C\to C\dg[\partial_\mu + \bb g\dg\partial_\mu  \bb g] C,
\ee
and motivates defining a gauge connection $\mathbb{ A}^C_\mu\equiv-i
\bb g^{-1}\partial_\mu \bb g$ which can be expanded $\bb
A^C_{\mu}=\frac{1}{2} \sum_a \Omega_{\mu} ^a\tau^a$ in terms of Pauli
matrices $\tau^a$, where $\Omega^{a}_{\mu}$ are the components of the
angular velocity associated with the Euler rotations. This gauge connection can be combined with the external
electromagnetic gauge potential as $\mathbb A=
\mathbb A^C-\tau^0
A^{ex}
$.
% Comment - this change is necessary to be consistent with p-eA for
% electromagnetism 
Setting $e=1$, the Lagrangian in imaginary time 
is 
\bea
{\cal L}&=&\bar C\Big[(\partial_\tau-\mu)\mathbb{1} +i\mathbb{A}_\tau-\frac{1}{2m_c}\sum_{\nu=1}^d (\partial_\nu\mathbb{1}+i\mathbb{A}_\nu)^2\Big]C\nonumber\\
&+&\bar f\Big[(\partial_\tau+\lambda)+iA^f_{\tau }+\frac{1}{2m_f}\sum_{\nu=1}^d(\partial_\nu+iA_\nu^f)^2\Big]f \nonumber\\
&+&
V[c\dg_1 f+ f\dg c_{1}]+i A^{ex}_\tau  n_{c}-iA^f_{\tau } Q.\label{eqfullL}
\eea
Here we have added two constraint terms, a term $i{A}^{ex}_\tau
n_{c}$ which account for the coupling of the fluctuations in the 
electromagnetic potential to the positive charge density of the ionic
background $n_{c}$, ensuring overall charge neutrality and the term
$iA^f_{\tau } Q$ which imposes the constraint $n_{f}=Q$ at each site. 
These terms ensure that when we carry out a gradient expansion, terms linear in
the gauge potentials vanish. 
From equation (\ref{eqfullL}), 
the action of the ungapped conduction electrons is given by
\bea
{\cal L}_{c2} &=& 
\bar c_{2}
\Big[
\partial_{\tau }+i
(A_{0} + {\textstyle \frac{1}{2}}\Omega^{3}_{\tau })\nonumber\\
&&\qquad\qquad+
\frac{1}{2m_{c}} \bigl [-i\nabla_{a}+ ({A}_{a}+{\textstyle
\frac{1}{2}}\Omega^{z}_{a})\bigr ]^{2}
\Big]c_{2}.\qquad
\eea
One of the interesting physical consequences of this action, is that
the propagation of the ungapped electrons in channel 2 picks up
the Berry phase associated with the spinor hybridization, so that
the vector potential acting on the ungapped electrons in channel 2 acquires  an additional component
associated with the Berry phase of the spinor field, 
\begin{equation}\label{}
A^{ext}_{\mu}\rightarrow A^{ext}_{\mu}+{\textstyle \frac{1}{2}}\Omega^{z}_{\mu}
\end{equation}
where $\Omega^{z}_{\mu}= -2 i z\dg  \partial_{\mu}z$ is the Berry
connection of the order parameter.

\subsection{Effective action}\label{effact}
The effective action for the gauge fields 
\be
Z=\int{[{\cal D}A^f {\cal D}A^{ex} {\cal D}A^{SU(2)}}]e^{-S_{\rm eff}}
\ee
can be obtained by integrating out the fermions (Appendix
\ref{NLSM}). 
A caricature of the long-wavelength action $S_{\rm eff}$
can be derived from a Landau-Ginzburg theory of the hybridization:
\begin{equation}\label{}
\frac{{\cal L}_{\rm 2CK}}{N}\sim  \frac{\hbar^{2}}{2m}\left|(-i\partial_{\mu}+
[A^{ext}_{\mu}-A^{f}_{\mu}])V\right|^{2}  + \frac{b}{2}\left(|V|^{2}-V_{0}^{2} \right)^{2} 
\end{equation}
The minimal coupling of the hybridization to the difference field
$A^{ext}_{\mu}-A^{f}_{\mu}$, is enforced by the gauge invariance
of the hybridization terms $(c\dg V f)+{\rm H.c}$. 
Under a
gauge transformation $c\rightarrow e^{i\theta }c$, $f\rightarrow e^{i\chi }f$,
$V\rightarrow e^{i (\theta -\chi )}V$, so that $V$ has the same
electrical charge as a conduction electron, but the opposite gauge charge to
an $f$-electron. 

At long distances, we may ignore amplitude fluctuations. Substituting
$V=V_{0}z$, where
\[
z  = {\mathbb g}\pmat{1\\
0 }=
e^{i \phi \tau_{3}/2
}
e^{i \theta \tau_{2}/2
}
e^{i \varphi \tau_{3}/2
}
\pmat{1\\
0 }.
\]
we obtain 
\bea
{{\cal S}_{\rm eff}}&=&\int{d^{d+1}x{\cal L}_{\rm 2CK}}\nonumber\\
{\cal L}_{\rm 2CK}&\sim &\frac{1}{2g}
\Big [
 \bigl(\partial_\mu \vec n\bigr)^2+ \bigl (\Omega_{\mu}^{z} - 2 [ A_\mu^{\rm ext}-A^f_\mu]\bigr )^2\Big].%\\
%&&+\Gamma_0^\tau(A_\tau^{\rm ext}-A^{z}_\tau)^2.
\label{eqL2CK}
\eea
where the implicit sum on $\mu\in [0,d]$ runs over all space-time dimensions.
Here, 
\be
\vec n= z\dg  \vec{ \tau }z= (\sin\theta\cos\phi,\sin\theta\sin\phi,\cos\theta),
\ee
is the channel magnetization, 
while 
\be
\Omega^z_\mu=[\partial_\mu\varphi+\partial_\mu\phi\cos\theta]=-i2 z\dg\partial_\mu z.\label{eqA3}
\ee
is the angular velocity of the spinor 
order about its principle $z$ axis and 
$g^{-1} = \frac{\hbar^{2}}{4m}V_{0}^{2}$.

Without the gauge coupling, this Lagrangian is 
the principal chiral field model, describing the evolution of a spinor
order parameter in space-time.\,\cite{Polyakov}
However, the coupling of the difference gauge field of 
$A_{\mu}^{ext}-A_{\mu}^{f}$ to rotations about the principle
``z'' axis of the order parameter,  ``Higgses'' the phase
fluctuations around the z-axis. 
The factor of two multiplying the coupling of 
the difference gauge fields 
reflects the fact that the channel magnetization as a vector,
carries integer channel quantum number 
whereas the electrons,
carry a half-integer channel quantum number, $\tau=1/2$. A detailed calculation of the effective action to one loop order in
the fermions is similar to the calculation of a superfluid stiffness
in a superconductor, and involves diagrams of the form depicted in Fig.\,\ref{fig:diagrams}(a).  These calculations (see Appendix \ref{NLSM})
confirm the basic form
obtained from the Landau-Ginzburg theory, but with  different
stiffnesses and mode velocities for rotations parallel and
perpendicular to the $\hat n$ axis.  The full Lagrangian, including the
contributions of the gapless electrons in channel 2 is then, 

\begin{widetext}

\bea
{\cal L}_{\rm 2CK}
&=&%\int d^{d}x  \left\{ 
\frac{N}{2g} 
\Big[\bigl(
\partial_{\tau }\vec n\bigr)^2 
+v_{g}^{2}\bigl(
\nabla\vec n\bigr)^2 
\Big]+ 
\frac{N\Gamma}{2}
\Big[
\Big(\Omega_{\tau}^{z} - 2 [ A_\tau ^{\rm ext}-A^f_\tau]\Big)^{2}
+ v_{\Gamma}^{2}
\sum_{i=1}^{d}\Big(\Omega_{i}^{z} - 2 [ A_i^{\rm ext}-A^f_i]\Big)^{2}
\Big]\nonumber\\
&&\qquad +
\bar c_{2}
\left[
\partial_{\tau }-i
[A_{0} + {\textstyle \frac{1}{2}}\Omega^{3}_{\tau }] +
\frac{1}{2m_{c}} \Big(-i\nabla_{a}- [{A}_{a}+{\textstyle
\frac{1}{2}}\Omega^{z}_{a}]\Big)^{2}
\right]c_{2}
%\right\}
%\\
%&&+\Gamma_0^\tau(A_\tau^{\rm ext}-A^{z}_\tau)^2.
\label{eqL2CK}
\eea

\end{widetext}

Note that the action scales as $N$, ensuring that the variance of the 
fluctuations about the large $N$ limit are of order $O (1/N)$. In the
limit that $m_{f}\gg m_{c}$, the  stiffness and velocity coefficients are given by (see Appendix C)
\begin{eqnarray}\label{eqcoefmain}
\frac{1}{g}& \approx & 2 \rho {\cal Z}, \qquad v_g\approx \frac{v_{c}}{\sqrt{\cal Z}}
\\
\Gamma&\approx&\frac{\rho}{4},\qquad \quad 
v_{\Gamma}\approx \sqrt{\frac{\pi^{2}}{2}v_{f}v_{c}}
\end{eqnarray}
Here $\tilde q_f=k_F^{2}/4\pi$ is the density of $f$-holes, $\rho={m_{c}}/{2\pi}$ is the conduction electron density of states, 
$v_{c}={k_{F}}/{m_{c}}$ and $v_{f}= {k_{F}}/{m_{f}}$ are the Fermi
velocities of the conduction and $f$-electrons respectively, while
\begin{equation}\label{}
{\cal Z}= \left[1+\frac{\rho T_K}{\tilde q_{f}}\Big(1+\frac{\tilde
q_f}{\rho_f T_K}\Big)^2\right] 
\end{equation}
is a mass renormalization factor, where $\rho_{f}= {m_{f}}/{2\pi}$  is
the f-electron density of states. 
 Note that 
in the limit where the Heisenberg coupling is zero  $J_{H}\rightarrow 0$, 
$m_f\to\infty$ and the axial stiffness
$(v_\Gamma)^{2}\Gamma\to 0$ associated with the Higgs term vanishes,
whereas the O(3)
stiffness associated with the channel magnetization $(v_g)^{2}/g$, 
remains finite.  

\subsection{The Anderson-Higgs term}

To understand the effect of the Anderson-Higgs term in \pref{eqL2CK} it is
useful to first consider the simpler
single-channel Kondo lattice\,\cite{Coleman05} where the effective action takes the form 
\be
{{\cal L}_{\rm 1CK}}=\sum_{\mu=0}^d
\frac{\Gamma}{2}
(\partial_\mu\varphi-2 (A^{ex}_\mu-A^f_\mu) )^2,\label{eqL1CK}
\ee
where $\Gamma=N/g$. 
The scaling dimension of this $\Gamma$ coupling is
$\dim[\Gamma ]=d-1$, making it relevant for $d>1$. For $d>1$, in the
ground-state, the internal $U(1)$ `vison' field $A^f$ is phase-locked to the
external gauge potential up to a pure gauge and a Meissner effect
develops for the difference field $A^{ex}_{\mu}-A^{f}_{\mu}$,
excluding the corresponding electromagnetic fields from the sample.
By fixing the gauge, we can  absorb the $\varphi$ field into the
vison gauge field $A^{f}$. 

Once the vison and electromagnetic fields lock, 
the conduction and $f$ electrons respond coherently to the common external
electromagnetic field, so the $f$-electrons acquire charge 
and now contribute to 
to the Fermi surface (FS) volume. 
Thus, the development of $f$-electron charge 
and the formation of a large FS in the Fermi liquid
regime of the Kondo lattice, as required by Oshikawa's
theorem\,\cite{Oshikawa} are all linked to this Anderson-Higgs effect. 
% Oshikawa considered a two-dimensional
%single-channel $SU(2)$ Kondo lattice on a torus (i.e. with periodic
%boundary conditions) and adiabatically threaded a flux quantum through
%the holes, after which the ground state returns to the original ground
%state up to a phase, which is associated with the total acquired
%momentum. By comparing this to the total momentum of a Fermi liquid,
%he showed that the spins contribute to the Fermi
%surface.\,\cite{Oshikawa}

A similar effect occurs in the two channel model, but now, the
absorption of the phase $\varphi$ into the gauge fields 
leaves behind the $(\phi,\theta)$ variables, which define the direction $\hat n (x)$ of
the channel magnetization.  One of the important distinctions here, is
that although the $\varphi $ field is Higgsed, the Berry phase term
of the spinor order is still present, described by the field 
\begin{equation}\label{}
\Omega^{z}_{\mu}\rightarrow \partial_{\mu}\phi \cos \theta 
\end{equation}
The survival of this term has important consequences, as we shall
shortly discuss. 

\begin{itemize}

\item the scaling behavior of the residual O(3) non-linear sigma model that
describes long-wavelength fluctuations of the channel magnetization,

\item the residual topological defects of the channel magnetization
field $\hat n (x)$.  These defects carry gauge charge. 

\end{itemize}

\subsection{The non-linear sigma model}\label{}

Once the Anderson-Higgs effect takes place, the residual 
long-wavelength behavior is described by an O(3) nonlinear sigma model
(NL$\sigma$M) with bare coupling constant $\tilde{g}= g/N$. 
In the large $N$ limit, the small size of $\tilde{ g}$ means that the
channel magnetization is always present in the ground-state. However,
we shall now consider the effects of scaling at finite $N$,
considering the stiffness $\tilde{g}$ to be finite.
The O(3)
$\sigma$-model  has been extensively studied in the
past.\,\cite{ZinnJustin,Polyakov,Sachdev} The coupling constant
$g$ 
has dimension $\dim[g]=2- (d+1)=1-d$ and its  renormalization flow at
weak coupling is determined by 
the beta function, 
\be
\beta ({g})=\frac{d{g}}{d\ell}=-\epsilon{g}+\frac{\Omega_{d+1}}{(2\pi)^{d+1}}{g}^2,
\ee
where $d\ell=-\log D$ and $\Omega_{d+1}$ is the solid angle in $d+1$ dimensions.
Above the lower critical dimension $d=1+\epsilon$, 
the scaling flow develops a new 
 fixed point at $ g_c\sim \epsilon$
corresponding to the quantum critical point between a disordered $g=\infty $
and ordered $g\rightarrow 0$ phase. 

At the lower critical dimension $d=1$, any value of the coupling constant
$g$ renormalizes to infinity $g\to \infty$ corresponding to a quantum
disordered (paramagnet) phase. One-dimensional two-channel Kondo lattices
have been studied in the past using bosonization\,\cite{Andrei00} as
well as density-matrix renormalization group\,\cite{Schauerte05} and
while there is evidence for non-Fermi liquid phases and
channel-antiferromagentic correlations, no channel-ferromagnetic state
was reported.

For $d>1$, a small bare coupling renormalizes to zero $g\to 0$,
corresponding to the ordered phase. However, for $g>g_c$, again the
system flows to the disordered phase $g\to\infty$. For $g<g_c$ and
$d>1$, the NL$\sigma$M describes the spontaneous breaking of
the channel symmetry and the consequent two Goldstone modes. They have
linear dispersion as the channel magnetization order parameter does
not commute with the Hamiltonian. The third Goldstone mode, associated
with the hybridization phase $e^{i\varphi}$ is Higgsed as we discussed before.

\subsection{Topological Defects}\label{}

The topology of the emergent channel magnetization admits skyrmions in
two dimensions and hedgehog defects in three dimensions ($\pi_{2} [O
(3)]={\mathbb{Z}}$).  Just as the Anderson-Higgs effect in a
superconductor
causes a vortex to bind 
a magnetic flux quantum in a superconductor, here, the 
the Berry phase of the skyrmion or hedgehog 
will bind a flux quantum in two dimensions, or a monopole 
in three dimensions. Both gapped, and ungapped electrons feel this field as a
\emph{physical} vortex or monopole field. 

For gapless $c_2$ electrons this is simply a consequence of the fact that they experience the gauge potential $A^{ex}_\mu-\frac{1}{2}\Omega_\mu$. We note that the curl of the Berry phase term is related to the curvature of
the $\hat n$ field via the Mermin-Ho relation, 
\begin{equation}\label{mermin}
\partial_{\mu}\Omega^{z}_{\nu }
-\partial_{\nu}\Omega^{z}_{\mu } = -\hat n \cdot (\partial_{\mu}\hat
n \times \partial_{\nu}\hat n).
\end{equation}
The quantity
\begin{equation}\label{}
\frac{1}{2}\int dS_{i}\epsilon_{ijk}\hat n \cdot (\partial_{j}\hat
n \times \partial_{k}\hat n) = 4\pi {\cal Q}
\end{equation}
measures the total solid angle swept out by the order parameter across
surface $S$, which is equal to $4\pi$ times the (integer) number ${\cal Q}$ of defects, hedgehogs
(3D) or skyrmions (2D), enclosed by the surface $S$.

Therefore, in absence of external potential, the phase accumulated by the $c_2$ electrons around  static defect is
\be
\frac{e}{\hbar}\Phi_{c2}=-\frac{1}{2}\int{d\vec S\cdot \nabla\times \vec\Omega^z}=-2\pi{\cal Q}
\ee
or $\Phi_{c2}={\cal Q}\frac{h}{e}$. Any transport experiment involves the gapless $c_2$ electrons and can potentially detect this experienced phase.

To understand how this works for the gapped electrons, note that 
the Higgs mass terms enforce the constraint 
\begin{equation}\label{higgsy}
\Omega^{z}_{\mu} = 2 (A^{ext}_{\mu}-A^{f}_{\mu})
\end{equation}
The connection between the Berry curvature and the vector potential fields
is closely analagous to a superconductor. In both cases, the
energetic requirement that supercurrents vanish at large distances
gives rise to the binding of flux to the defect.

\begin{figure}[htb]
{\includegraphics[width=\columnwidth]{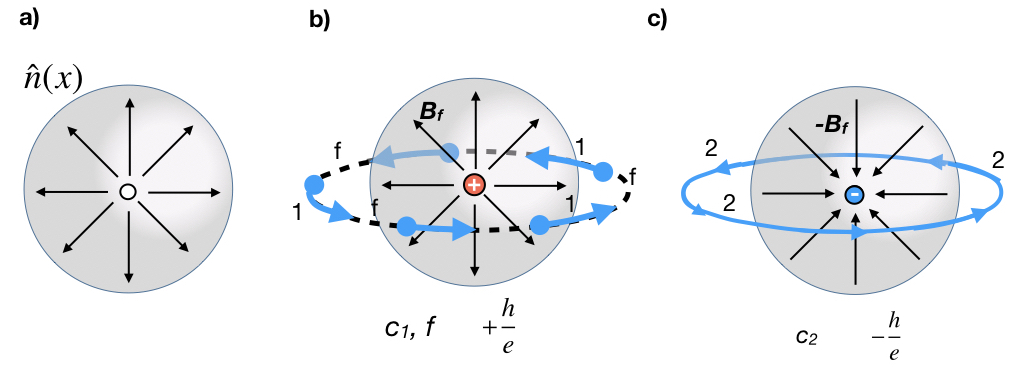}}
\caption{\raggedright 
(a) Hedgehog defect of the $\hat n$ field in 3 dimensios. 
(b) Hybridized conduction
and $f$-fermions in channel 1 
see the defect as a positive monopole (b) unhybridized
$c$-fermions see the defect as a negative monopole. 
}\label{weiss}
\end{figure}

 Using the Mermin-Ho relation (\ref{mermin}) and the Higgs constraint
(\ref{higgsy}), it follows that
\begin{eqnarray}\label{l}
4 \pi {\cal Q} &=& -\int dS_{i}\epsilon_{ijk}\partial_{j}\Omega_{k} 
\cr
&=&  -2
\int d\vec{ S}\cdot \vec{\nabla}\times (\vec{A}^{ext}-\vec{A}^{f})
\end{eqnarray}
or
\begin{equation}\label{}
\int d\vec{ S}\cdot \vec{\nabla}\times (\vec{A}^{f}-\vec{A}^{ext})
= \int d\vec{S}\cdot (\vec{B}^{f}- \vec{B}^{ext})
= 2 \pi Q.
\end{equation}
This tells us that the topological defect must bind a flux quantum of
the difference field,  much in the way a superconducting vortex binds a
magnetic flux quantum.  In the case of a skyrmion,  this corresponds to a
magnetic flux quantum of the difference field. (In practice the larger
energy cost of a magnetic field will likely mean the bound quantum is
largely in the f-field). 
However, for a hedgehog defect in 
three dimensions the strict absence of electromagnetic monopoles
guarantees that $\int d\vec{S}\cdot \vec{B}^{ext}= \int d^{3}x
\vec\nabla \cdot \vec{B}^{ext}=0$.  In this case we can eliminate the
electromagnetic
component of the surface integral. 
This means that the total f-flux bound to a ${\cal Q}$
hedgehog is \begin{eqnarray}\label{l}
\frac{e}{\hbar }\Phi_{f} =  \int d\vec{S}\cdot \vec{\nabla }\times
\vec{A}^{f} = 2 \pi {\cal Q}
\end{eqnarray}
where we have restored the $e/\hbar $ to the definition of flux. Thus
\begin{equation}\label{}
\Phi_{f} = {\cal Q} \frac{h}{e}
\end{equation}
so that each hedgehog or skyrmion carries unit magnetic
flux of the $f$-field, forming a vison monopole. 

Remarkably, when we work it through, we find that the Berry phase
field experienced by the conduction electrons means that they also feel
the vison field.   In fact, 
the $f$-electrons and the
conduction electrons they hybridize with in channel 1 experience
the vector potential $\vec{A}^{f}$, whereas the gapless  channel 2
fermions experience a vector potential $-\vec{A}^{f}$, as if the two
channels have acquired an opposite charge. 
To see this,
let us first set the electromagnetic potential to zero
$A^{ext}_{\mu}=0$.  The resulting vector potential of the conduction electrons
is then 
\begin{equation}\label{}
\mathbf{A}_{a}= -\frac{1}{2}\Omega^{z}_{a}\tau_{3}\rightarrow
A_{f}\tau_{3}\equiv  \left\{
\begin{array}{rl}
A_{f}&\qquad c_{1}\cr
-A_{f}&\qquad c_{2}
\end{array}
 \right.
\end{equation}
In this way all electrons feel the 
monopole, in such a
way that the hybridized and unhybridized electrons experience an
equal and opposite monopole vector potential.  
One of the interesting effects of the monopole field will be 
to produce bound-states in the gap. 
Within the ordered phases, this opens up the interesting possibility
that topological configurations
of the order parameter can be detected by purely electronic transport
studies.

\subsection{Phase diagram} In the channel ferromagnetic phase the
skyrmions are gapped,\,\cite{Belavin75} with an energy given by
$4\pi{\cal Q}g^{-1}$. Therefore, at low-enough temperature, the action
\pref{eqL2CK} reduces to a separate sum of a $U(1)$ field, as in
a single-channel Kondo insulator, Eq.\,\pref{eqL1CK}, and the
NL$\sigma$M term describing the fluctuations of the Goldstone
modes. In this phase, we have the phase-locking $A^f=A^{ex}$ and as
discussed above, a large total Fermi surface (albeit only the Fermi
surface of one conduction band is expanded).

Close to a quantum critical point into another ordered phase, e.g,
magnetism, the Kondo temperature is suppressed to
zero\,\cite{Komijani18a,Komijani18b} and the Higgs mode also becomes
soft. The correct description then, includes this soft term and is
beyond the treatment adapted here.

\begin{figure}[t!]
\includegraphics[width=\linewidth]{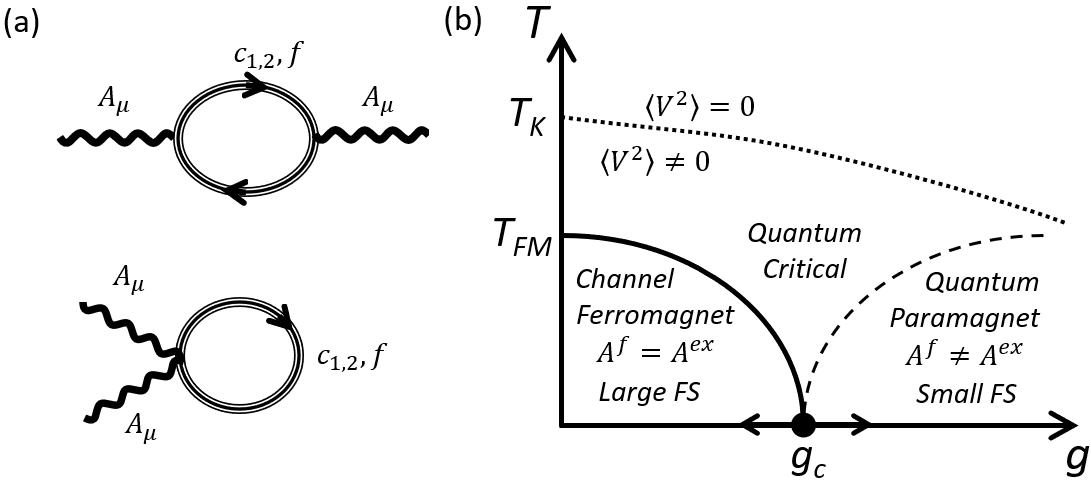}
\caption{\raggedright\small (a) The basic diagrams (paramagnetic and diamagnetic contributions) involved in integrating out the fermions to one loop. (b) The phase diagram for $d=3$. The same phase diagram applies to $d=2$, except that $T_{FM}\to 0$. The sold lines are second order phase transition and the dashed line is a crossover. In the channel ferromagnet $g\to 0$. In this phase internal and external gauge fields are phase-locked $A^f=A^{ex}$, resulting in a large Fermi Surface (FS). For $g>g_c$ the system is in the quantum disordered phase, where $A^f\neq A^{ex}$ and the Fermi surface is smaller.}
\label{fig:diagrams}
\end{figure}

However, within the $\langle{V^2}\rangle\neq 0$ regime, we expect a separate
quantum critical point defined by the NL$\sigma$M physics at $g=g_c$. For $g_0>g_c$ within
perturbative RG, $g\to\infty$ and the ground state is disordered. The
nature of the groundstate in this phase is quite interesting and we
can gain a guide to it using the physics of the 
the  $O(M)$ NL$\sigma$M. In the large $M$ limit of this model, we know that 
the $z$-spinons (long-wavelength fluctuations of channel
magnetization $\vec n$) are gapped.\,\cite{Sachdev} However, the
topological defects proliferate and
condense.\,\cite{Polyakov,Read90,Senthil04} This process will
eliminate the phase-locking between the electromagnetic and $f$-fermion 
gauge fields, and the resulting electronic Fermi surface will become
small again. This is schematically shown in
Fig.\,\ref{fig:diagrams}(b). Were such a phase transition to occur in
a real material, we might expect it to exhibit a 
jump in the FS. 
In contrast to a magnetic transition, the
magnetic susceptibility is expected to remain finite at this
transition.

Besides having a small Fermi surface, the nature of the ground state in
the quantum disordered phase remains unclear. Assuming that the
resulting phase is a Fermi liquid, this small FS appears to violate
 Oshikawa's theorem.\,\cite{Oshikawa} A trivial resolution to this
paradox might be that the adiabatic assumption of the flux-threading is
violated due to the gapless nature of the spin-liquid. However, as
discussed before, the hybridization is non-zero in the quantum
disordered phase and the $f$-spinons are likely to remain gapped, as
in the channel ferromagnet phase. Therefore, a better resolution to
this paradox is that the ground state has topological order
(i.e. degeneracy on a torus).\,\cite{Senthil03} In that case, after an
adiabatic threading of a flux through the torus holes the system need
not be back to the same ground state. Based on this, we conjecture
that the quantum disordered phase of a two-channel Kondo lattice may
have topological order, realizing a fractionalized Fermi liquid
(FL$^*$).

%The change in the size of the Fermi surface is a consequence of the coupling between NL$\sigma$M and the Higgs term and clearly goes beyond conventional Landau-Ginzburg paradigm. On the other hand, the deconfined critiality\,\cite{Senthil04} devised to capture certain phase transitions in antiferromagnets involves an emergent fractionalized $z$-spinons that only exists at the quantum critical point and in neither of the two phases. What is new here, is the extension of the deconfinement to an entire phase (order fractionalization), a phenomena we discussed in the previous section.\,\cite{Komijani18}

\section{Conclusion and Outlook}\label{conclusion} 

We have shown that in the large-$N$ limit
the ground state develops a spontaneously broken channel symmetry, 
most naturally understood in terms of order fractionalization:
a process which involves the
separation of the composite spin-fermion bound states into 
a fermionic resonance and a half-integer order parameter, 
manifested in the long-time behavior of the electronic self-energy.

Our analysis of collective soft modes shows that the effective action
is composed of a non-linear sigma term describing symmetry breaking
and the Goldstone modes, and a Kondo-Higgs term which causes the phase-locking of internal and external gauge fields, and the
expansion of the Fermi surface. These two terms become 
intertwined in the presence of topological defects, which behave as
monopoles 
with a  $U(1)$ gauge charge which locally
destroys the phase locking between the fields. This
allows us to predict that when these defects proliferate in the
quantum disordered phase of the two-channel Kondo lattice, the Fermi
surface jumps from large to small, with experimental consequences.

These arguments lead us to expect that 
in addition to the channel ferromagnetic phase, the higher-dimensional
(two or three-dimensional)
two-channel Kondo lattices, 
have a quantum disordered phase, possibly with topological order. 
Although this phase has not yet been seen in 
experiments using pressure or magnetic field as the tuning
parameter, it may be revealed by using these tuning parameters in
combination
to allow a more extensive exploration of the phase diagram.

 Our results suggest a number of interesting directions for future
work.  For example, a more complete analysis of the
particle-hole symmetric two-channel Kondo soft modes will need to take into account the full
$SP(4)\sim SO(5)$ symmetry of the problem,\,\cite{Affleck92} a symmetry  that allows the
rotation between channel magnets and composite paired (odd-frequency) superconducting
ground-states. At the impurity level (or in
quantum disordered phases in larger $d$), the order parameter strongly
fluctuates, exploring the full symmetry group. This appears to be
responsible for capturing the residual entropy of the two-channel
Kondo impurity. Moreover, the term $C\dg A^z_\tau\tau^z C$ in the
Lagrangian \pref{eqfullL} can be interpreted as a Berry phase for the
order parameter $\vec n$. Within one-loop and at half-filling, the coefficient of this
term $\langle C\dg\tau^zC\rangle$ is zero. However, $C\dg\tau^zC$ does
not commute with the Hamiltonian and the Berry phase might have
important effects beyond one-loop.

Further work is also required to understand the ground state and
excitations of the quantum-disordered phase of the two channel Kondo
lattice and possible topological order that may develop in this phase.
In particularly, the relationship of this phase to deconfined
criticality will require studying defect proliferation.

\begin{acknowledgments}
It is a pleasure to thank Premala Chandra, Indranil Paul, Elio K\"onig, Weida Wu, Senthil Todadri, Philippe Gegenwart and Achim Rosch for fruitful discussions. P.~C. and A.~W. were supported by the National Science Foundation grant
DMR-1830707. Y.~K. was supported by a Rutgers University Materials Theory postdoctoral fellowhsip.
\end{acknowledgments}

\section*{Appendices}
The appendices contain additional proofs and details that are used in the paper. The uniform mean-field solution is compared to an alternative staggered solution in Appendix \ref{afm}. In Appendix \ref{continuum} we describe a continuum model for the Kondo lattice that is used in field theory calculations. Appendix \ref{NLSM} contains a derivation of the effective action, which is a central point of the paper.
\appendix
\section{Staggered Hybridization Solution}\label{afm}

\begin{figure}[t!]
\includegraphics[width=\linewidth]{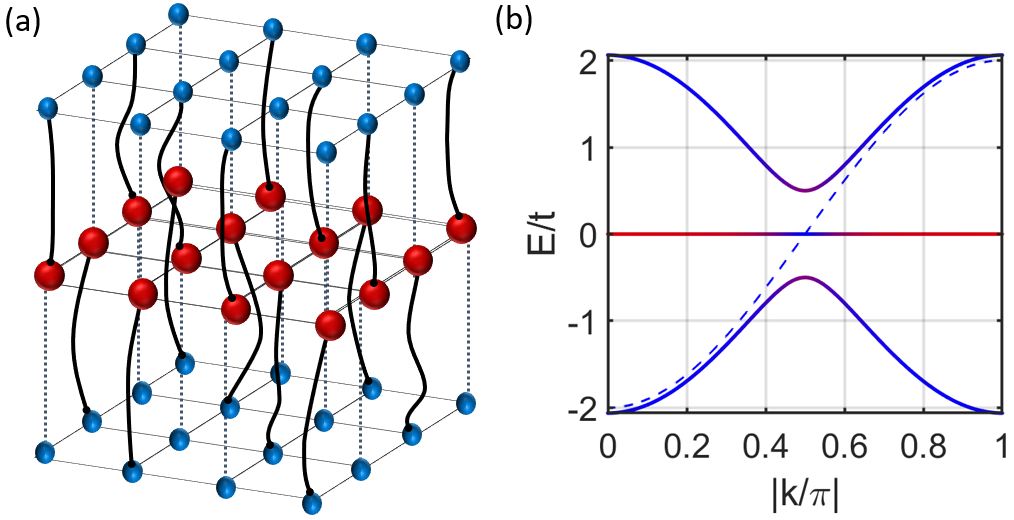}
\caption{\small\raggedright (a) Channel antiferromagnetic state: the hybridization between the local moments (red) alternates between the two channels at every other site. (b) Schematic representation of the band structure of the staggered hybridization state.}
   \label{fig:bandStructStagHyb}
    \label{fig:mainlabStagHyb}
\end{figure}

In this section, we compare the half-Kondo-insulator solution with an alternative channel symmetry breaking mean-field state, for which the hybridization is staggered in channel space. The corresponding mean field configuration is  
\bea
&&\lambda_j =0, \quad V_j=V, \qquad\varphi_j=\varphi,\\
&& \phi_j=0,\qquad \theta_j=\frac{\pi}{4}\left[1-e^{i\pi( j_1 + j_2)}\right].  
\eea
The hybridization term can be further simplified by choosing the gauge with
\begin{equation}
\begin{pmatrix}\tilde{c}_{\vec{k}1\alpha} \\
\tilde{c}_{\vec{k}1\alpha} \\
\end{pmatrix} = \frac{1}{\sqrt{2}}\begin{pmatrix}1 & 1 \\
1 & -1 \\
\end{pmatrix}\begin{pmatrix}c_{\vec{k}1\alpha} \\
c_{\vec{k}2\alpha} \\
\end{pmatrix},\\
\tilde{f}_{\vec{k}\alpha}=e^{i\phi}f_{\vec{k}\alpha.}
\end{equation} 
The band energies are each doubly degenerate with values 
\begin{equation}
E^{\mathrm{1,2}}=0,\ E^{\mathrm{3,4}}_{\vec{k}}=-\sqrt{\epsilon_{\vec{k}}^2+V^2},\ E^{\mathrm{5,6}}_{\vec{k}}=\sqrt{\epsilon_{\vec{k}}^2+V^2}.
\end{equation}

The free energy per site per particle is then given by

\bea
\frac{F}{N{\cal N}_s}&=&-4T\int\limits_{\frac{1}{2}\mathrm{B.Z.}}\frac{\mathrm{d}^2k}{(2\pi)^2} \log\left[2\cosh\left(\frac{\sqrt{\epsilon_{\vec{k}}^2+V^2}}{2T}\right)\right]\nonumber\\
&&\qquad-2T\log(2)+\frac{V^2}{J},
\eea
and the hybridization strength is determined self-consistently from
\be
\frac{1}{N{\cal N}_s}\frac{\delta F}{\delta V^2}=\frac{1}{J}-\int\limits_{\frac{1}{2} \mathrm{B.Z.}}\frac{d^2 k}{(2\pi)^2} \frac{\tanh\left(\frac{\sqrt{\epsilon_{\vec{k}}^2+V^2}}{2T}\right)}{\sqrt{\epsilon_{\vec{k}}^2+V^2}}=0.\nonumber
\ee
Using a constant density of states $\rho=(8t)^{-1}$, we can estimate the hybridization strength at zero temperature to be
\begin{equation}
V^{-1}={2\rho\sinh\left(\frac{1}{\rho_0 J}\right)}.
\end{equation}

%\begin{figure}[h!]
%    \centering    
%    \includegraphics[width=0.5\textwidth]{gsEnerStagHyb}
%    \caption{Ground state energy of the staggered hybridization state}
%    \label{fig:gsEnerStagHyb}
%\end{figure}

In figure \ref{fig:energy}(b), the free energy is plotted for varying hybridization strength for both uniform and staggered hybridization. As stated before, the uniform solution has lower energy and is therefore a better candidate for the ground state.

\section{A Continuum Kondo Insulator}\label{continuum}
Kondo lattices in the large-$N$ limit are usually studied on tight-binding models. For the derivation of the NL$\sigma$M and the study of the skyrmion spectrum it is much easier to use an approximate low-energy description of the Kondo lattice which has continuous translational and rotational symmetry. In other words, we consider the momentum-space Hamiltonian
\be
\hspace{-.15cm}H=\sum_k\mat{c\dg_k & f\dg_k}\mat{\eps_k^c & V \\ V &
\eps_k^f}\mat{c_k \\ f_k}+N{\cal N}_{s}
\left(\frac{V^2}{J_K}-\lambda q_{f} \right)
\label{eqcki}
\ee
where
\be
\eps_k^c=\frac{k^2}{2m_c}-\mu, \quad\text{and}\quad \eps_k^f=-\frac{k^2}{2m_f}+\lambda,
\ee
and $q_{f}=Q_{f}/N$.
Eq.\,\pref{eqcki} describes the low-energy Hamiltonian of a Kondo-Heisenberg system. The dispersion of $f$-electrons arises due to antiferromagnetic Heisenberg  interactions between the spins. In order to have a Kondo insulator we have assumed opposite sign of mass for conduction and $f$-electrons. Due to lack of particle-hole symmetry, it is not clear whether a continuum version of a Kondo insulator exists. The main challenge is to show that the conditions of having a spectral gap and the constraint $n_f=Q_{f}$ can be simultaneously realized in this system. To be specific, we limit our discussion to $d=2$ spatial dimensions. 

The Hamiltonian\,\pref{eqcki} can be diagonalized using an O(2) rotation
\be
\mat{c_k\\ f_k}=\mat{\cos\alpha_k & -\sin\alpha_k \\ \sin\alpha_k & \cos\alpha_k}\mat{l_k \\ h_k}
\ee
where
\be
\tan2\alpha_k=\frac{2V}{\eps_k^c-\eps_k^f}
\ee
leading to the energy eigenvalues\be
E^{l/h}_k=\frac{\eps_k^c+\eps_k^f}{2}\pm\sqrt{\left(\frac{\eps_k^c-\eps_k^f}{2}\right)^2+V^2}.
\ee
Due to $\pi$-periodicity of the $\tan 2\alpha_k$, we are free to choose either the period $2\alpha_k\in (0,\pi)$ or $2\alpha_k\in (-\pi/2,\pi/2)$. We choose the former interval, because the angle evolves more continuously in the Brillouin zone. Therefore,
\be
\sin2\alpha_k=\frac{2V}{E_k^l-E_k^h}, \qquad 
\cos2\alpha_k=\frac{\eps_k^c-\eps_k^f}{E_k^l-E_k^h}.
\ee
For a Kondo insulator, the $E^h$ band is fully occupied, while the
$E^l$ band is empty, so that the ground-state free energy is
\be
\frac{F}{N{\cal N}_{s}}=\frac{V^2}{J_K}+\int{\frac{d^2k}{(2\pi)^2}E_k^h}-\lambda q_{f}
\ee
Varying $F$ with respect to $V^2$ gives the  mean-field equation
\bea
\frac{1}{J_K}&=&\int{\frac{d^2k}{(2\pi)^2}}\frac{1}{\sqrt{(\eps_k^c-\eps_k^f)^2+4V^2}}\label{eqB62}
\eea
while varying it with respect to $\lambda$ enforces the constraint.
For the 
inverted $f$-electron band, it is more convenient to apply the
constraint to the f-hole occupation 
\bea
\tilde q_f\equiv 1-q_f&=&\frac{1}{L}\sum_k\braket{f\dn_k f\dg_k}.
\eea
In the absence of hybridization in $d=2$, the f-hole density is
\be
\tilde q_f=\int{\frac{d^2k}{(2\pi)^2}} f(-\eps^f_k)\xrightarrow{T=0}\frac{{m_f}}{2\pi}\abs{\lambda_0}
\ee
Similarly, the density of electrons (per spin) is given by 
\be
\qquad q_{c}=\frac{m_c}{2\pi}\mu.
\ee
In presence of hybridization and temperature much lower than the gap, only the $E^h$ band is occupied. Therefore, the density of $f$-holes and $c$-electrons is the same and is given by
\bea
\tilde q_f=q_{c}&=&\int{\frac{d^dk}{(2\pi)^d}}\sin^2\alpha_k=\nonumber\\
&=&\frac{1}{2}\int{\frac{d^dk}{(2\pi)^d}}\left[1-\frac{\eps_k^c-\eps_k^f}{\sqrt{(\eps_k^c-\eps_k^f)^2+4V^2}}\right]\qquad\label{eqqf2}
\eea
In the continuum limit, eqs.\,\pref{eqB62} and \pref{eqqf2} become
\bea
\frac{1}{J_K}&\approx &\nu\sinh^{-1}\Big[\frac{\Lambda^2}{8\pi\nu V}\Big(\eta+\sqrt{\eta^2+1}\Big)\Big]\label{eq2B13}\\
\tilde q_f&\approx &\nu V\sqrt{\eta^2+1}\label{eq2B14}
\eea
where $\nu \equiv [{{2\pi}/{m_c}+{2\pi}/{m_f}}]^{-1}$ is an average density of states of the $c$ and $f$ bands,  $\Lambda$ is the high-energy momentum cut-off and $\eta\equiv (\mu +\lambda)/2V$ is a short-hand notation.  The first equation suggests defining
\be
T_K\equiv  \frac{\Lambda^2}{8\pi\nu\sinh\left(\frac{1}{\nu J_K}\right)}.
\ee
In terms of $T_{K}$, eqs.\,(\ref{eq2B13}) and (\ref{eq2B14}) can be solved for
\begin{align}
V=\sqrt{2 D T_K -T_K^2}\ \ \mathrm{and}\ \ \lambda\approx
 2D-\mu-2T_K,
\end{align}
where $D\equiv {\tilde{q}}_f/{\nu}\gg T_K$ is an emergent energy
scale, which plays the role of an effective bandwidth and we have
expanded the expression for $\lambda$ to leading order in  ${T_K}/{D}$. In the following, it is convenient to introduce two dimensionless parameters:
\be
t_K\equiv T_K/D, \qquad r_m\equiv m_c/{m_f}
\ee
in terms of which
\be
V/D\approx ({\eta^2+1})^{-1/2}\approx \sqrt{2t_K}.\label{eq2B18}
\ee
For a physical realization of a Kondo insulator we expect $t_K\ll 1$ and $r_m\ll 1 $ and all following expressions assume this limit. The presence of a Kondo insulator, requires having the chemical potential inside the gap. In particular, the minimum of the upper band, $E^l$, has to be above zero energy and the maximum of the lower band, $E^h$, has to be below zero energy. This constraints the chemical potential to be in the region  
\begin{align}
2 -2t_K-\sqrt{8r_m t_K}\lessapprox\mu/D\lessapprox 2-2t_K,
\end{align}
for $t_K \gtrapprox 2r_m$ and
\begin{align}
\hspace{-.5cm}2-2t_K-\sqrt{8r_m t_K}\lessapprox\mu/D\lessapprox2-2t_K+\sqrt{8r_mt_K}
\end{align}
for $t_K \lessapprox 2r_m$. The gap closes in either case of $r_m\to 0$ or $t_K\to 0$. However, the order of limits matters and the indirect gap is given by $D\sqrt{8r_mt_K}$ and $2D\sqrt{8r_mt_K}$ for $t_K>2r_m$ and $t_K<2r_m$, respectively.

\begin{figure}[t]
\centering
\includegraphics[width=0.45\textwidth]{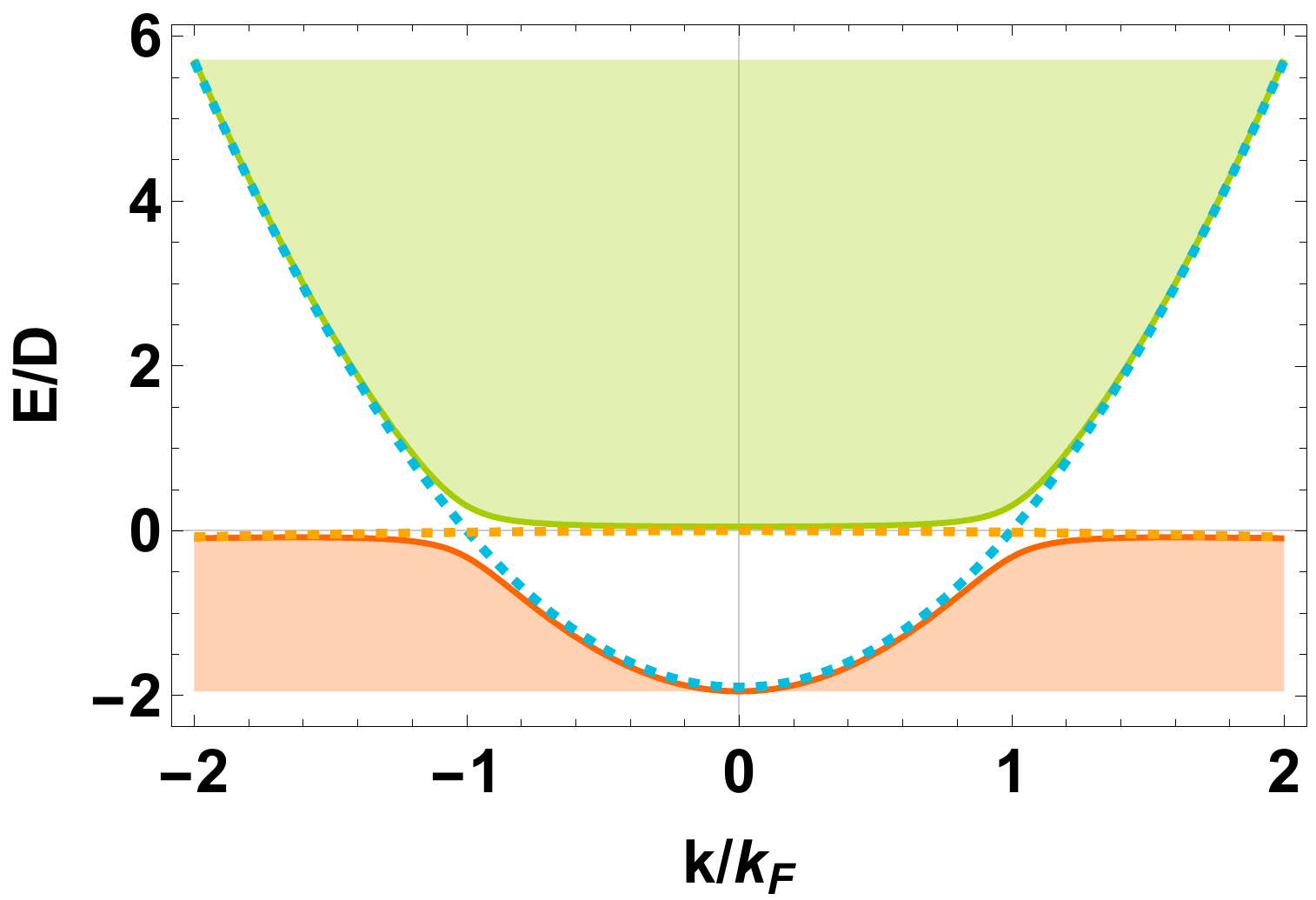}
\caption{\small\raggedright An example of a continuum Kondo insulator in 2D. The model parameters are ${m_f}/{m_c}=100$, ${\mu}/{D}=1.9$ and ${T_K}/{D}=0.05$. The dashed green line corresponds to the $c$-electrons, while the $f$-electrons are depicted by the dashed orange line. After these two bands hybridize, we obtain a lower band, $E^h$ (solid orange  line), and an upper band, $E^l$ (solid green line).   \label{fig:contKondoInsBandStruct}}
\end{figure}

An example of a band structure of such a continuum Kondo insulator is depicted in Fig.\,\pref{fig:contKondoInsBandStruct}.

\section{Derivation of the Effective action}\label{NLSM}
We start from the Lagrangian \pref{eqfullL}. Integrating out the fermions, the free energy is
\bea
F(A)=-NT\tr\log[-{\cal G}^{-1}(A)]-F_{b} (A) ,
\eea
where
\be
-{\cal G}^{-1}(A)=\matc{c|c}{-\bb g^{-1}_C & \matl{V \\ 0} \\ \hline
\matl{V & 0} & -g^{-1}_f},
\ee
and
\begin{equation}\label{}
F_{b}[A]=i (A^{ex}_\tau
n_{c}-A^f_{\tau } Q){{\cal N}_{s}}
,
\end{equation}
is the background free energy due to the 
 coupling of the
potentials to the background conduction and f-electron charge.
These terms
ensure that when we expand the effective action in the applied fields,
terms linear in the coupling identically vanish. 
We can write
\begin{eqnarray}\label{l}
\Delta F(A)&\equiv &F(A)-F(A=0)\cr
&=&-NT\tr\log\left\{
1-{\cal G}{\cal W}\right\}-F_{b} (A)
\end{eqnarray}

where ${\cal G}\equiv{\cal G}(A=0)$ and ${\cal W}$ are given by
\be
{\cal W}=\matc{c|c}{\bb W & \matl{0\\0} \\ \hline \matl{0 & 0 }& w^f}, \quad {\cal G}=\matc{c|c}{\bb G_C & \matl{G_{1f}\\ 0} \\ \hline \matl{G_{1f} & 0} & G_f},
\ee
where in the momentum-frequency domain the Green's functions are given by \begin{gather}
\mathbb{G_C} = \mathrm{diag}\left[G_1=\frac{z-\varepsilon^f_k}{\left(z-E^l_k\right)\left(z-E^h_k\right)}, g_2= \frac{1}{z-e^c_k}\right]\\
G_{1f}=\frac{V}{\left(z-E^l_k\right)\left(z-E^h_k\right)}\\ G_f=\frac{z-\varepsilon^c_k}{\left(z-E^l_k\right)\left(z-E^h_k\right)}.
\end{gather}
In order to take the trace over channel indices, it is convenient to expand $\mathbb{G_C}$ in terms of Pauli matrices in channel space $\mathbb{G_C}=G^0\tau^0+G^3\tau^3$.
Additionally, we have defined the short-hand symbols
\bea
\bb W&=&i\bb A_\tau-\frac{1}{2m_c}\sum_{\nu=1}^d[i\partial_\nu\bb A_\nu+2i\bb A_\nu\partial_\nu-\bb A_\nu\bb A_\nu]\nonumber\\
w^f&=&iA^f_\tau+\frac{1}{2m_f}\sum_{\nu=1}^d[i\partial_\nu A^f_\nu+2i A^f_\nu\partial_\nu-A^f_\nu A^f_\nu],\qquad
\eea
where, following section (\ref{softmodes}), the conduction electron 
vector potential contains a Berry phase and an electromagnetic term, given by ${\bb A}_{\mu}= A^{a}_{\mu}\tau_{a} =
{\frac{1}{2}}\Omega^{a}_{\mu}\tau_{a}- \tau_{0}A^{ext}_{\mu}$.
Defining
\be
{\cal X}={\cal G}{\cal W}=\matc{c|c}{\bb G_C\bb W & \matl{G_{1f}w^f \\ 0} \\ \hline \matl{G_{1f}\left[\bb W\right]_{11} & 0} & G_fw^f}
\ee
and expanding the log 
\be
-\log(1-{\cal X})\approx {\cal X}+{\cal X}^2/2,
\ee
to $O(A^2)$ leads to

\bw

\bea
\frac{\Delta F}{NT}={\rm Tr}&\Big\{&\bb G_c\bb Q+G_fq_f+\frac{1}{2}(\bb G_C\bb Q)^2+\frac{1}{2}(G_fq_f)^2+G_{1f}Q_{11}G_{1f}q_f\Big\}-F_{b}[A]\\
=
\tr&\Big\{&\frac{1}{2m_c}\bb G_C \bb A_\nu\bb A_\nu
+\frac{1}{2m_f}G_fA_\nu^fA_\nu^f
%\nonumber\\&&
-\frac{1}{2}\bb G_C\bb A_\tau \bb G_C \bb A_\tau
-\frac{1}{8m_c^2}\bb G_C\Big(\cancel{\partial_\nu \bb A_\nu}+2\bb A_\nu\partial_\nu\Big)\bb G_C\Big(\cancel{\partial_{\nu'}\bb A_{\nu'}}+2\bb A_{\nu'}\partial_{\nu'}\Big)\nonumber\\
&&-\frac{1}{2} A_\tau^fG_fA_\tau^fG_f
-\frac{1}{8m_f^2}(\cancel{\partial_\nu A_\nu^f}
+2A_\nu^f\partial_\nu)G_f(\cancel{\partial_{\nu'}A_{\nu'}^f}+2A_{\nu'}^f\partial_{\nu'})G_f
%\nonumber\\&&
-G_{1f}A_{\tau}^{C,11}G_{1f}A^f_\tau
\nonumber\\
&&\frac{G_{1f}}{4m_cm_f}(\cancel{\partial_\nu A_\nu^{C,11}}+2A_\nu^{C,11}\partial_\nu)G_{1f}(\cancel{\partial_{\nu'}A_{\nu'}^f}+2A_{\nu'}^f\partial_{\nu'})\Big\}.
\eea
\ew

Here, the trace is taken over all space/time and channel variables.
The terms linear in the applied fields vanish, because the net charge
densities and currents are identically zero in the ground-state, while
the terms containing odd numbers of time or space  derivatives also
vanish, as they are odd under time-reversal or spatial inversion. 
We have also omitted higher derivatives of $A$ (see below). Lastly, anticipating the long-wavelength limit ($\vec{q}\rightarrow 0$), we have neglected the terms that contain the divergence of the gauge fields.

\subsection{Terms quadratic in $A_\tau$}

In momentum space, the quadratic terms have the generic form 
\bea
&&\frac{1}{\beta}\sum_{i\omega_n,i\nu_r}\int\frac{\mathrm{d^2}k\mathrm{d^2}q}{(2\pi)^4}G(i\omega_n,\vec{k})A(i\nu_r,\vec{q})\nonumber\\
&&\hspace{2.5cm}G(i\omega_n+i\nu_r,\vec{k}+\vec{q})A(-i\nu_r,-\vec{q}).\qquad
\eea
Here, we assume a slow variation of the gauge potentials in space/time and only keep the lowest order in $i\nu_r$ and $\vec{q}$ in the Green's functions.  Terms quadratic in gauge potential and containing only the time-components are
\bea
\frac{\Delta F_{temporal}}{NT}&=&{\Lambda^{00}}\int{d\bar x_\mu} \sum_{a=0}^3[A^a_\tau(\bar x_\mu)]^2\nonumber\\
&+&{\Lambda^{33}}\int{d\bar x_\mu}\Big\{\sum_{a=0,3}-\sum_{a=1,2} \Big\}[A^a_\tau(\bar x_\mu)]^2\nonumber\\
&+&{4\Lambda^{03}}\int{d\bar x_\mu}A^0_\tau (\bar x_\mu)A_\tau^3(\bar x_\mu)\nonumber\\
&+&\frac{\Lambda^{ff}}{2}\int{d\bar x_\mu}[A_\tau^f(\bar x_\mu)]^2\nonumber\\
&+&\Lambda^{1f,1f}\int d{\bar x_\mu}[A_\tau^0
(\bar x_\mu)+A_\tau^3(\bar x_\mu)]A_\tau^f
(\bar x_\mu).\qquad
\eea
where we have defined
\bea
\Lambda^{ab}&\equiv&-\lim_{i\nu_r\rightarrow 0,\vec{q}\rightarrow 0}\frac{1}{\beta}\sum_{i\omega_n}\int\frac{\mathrm{d^2}k}{(2\pi)^2}\nonumber\\
&&\hspace{2cm}G^a(i\omega_n,\vec{k})G^b(i\omega_n+i\nu_r,\vec{k}+\vec{q}),\qquad\quad
\eea
where the $G^{a}$ are the components in the Pauli-matrix decomposition
$\mathbb{G_C}=G^0\tau^0+G^3\tau^3$ of conduction electron propagator. 
The order of limits as indicated is crucial for extracting gauge-invariant results. These terms are of the form
\bea
&&(\Lambda^{00}+\Lambda^{33})[(A^0_\tau)^2+(A_\tau^3)^2]+(\Lambda^
{00}-\Lambda^{33})[(A^1_\tau)^2+(A_\tau^2)^2]\nonumber\\
&&+4\Lambda^{03}A_\tau^0A_\tau^3+\frac{1}{2}\Lambda^{ff}(A_\tau^f)^2+\Lambda^{1f,1f}[A_\tau^0+A_\tau^3]A_\tau^f.\qquad
\eea
We can compute the coefficients using mean-field Green's functions. We find
\begin{gather}
\Lambda^{00}+\Lambda^{33} = 2\Lambda^{03}=\Lambda^{ff}=\frac{1}{2}\Lambda^{1f,1f}.
\end{gather}
as demanded by the gauge invariance of the original Hamiltonian. Various terms can be combined and the effective Lagrangian contains
\begin{gather}
\frac{{\cal L}_{temporal}}{N}=2\Gamma(A_\tau^0+A_\tau^3-A_\tau^f)^2+\frac{1}{2g}[(A_\tau^1)^2+(A_\tau^2)^2].
\end{gather}
Here the 
definition of parameters is deliberate as we recognize the Higgs term from 
section \ref{effact}. In appendix \ref{altrep} we will identify the
second term 
as the temporal part of the NL$\sigma$M.\\ 

We can calculate the coefficients explicitly in the limit $T\ll T_K$ for the continuum model discussed in the previous section.
\bw

\begin{gather}
2\Gamma=\Lambda^{00}+\Lambda^{33}=-\lim_{i\nu_r\rightarrow 0,\vec{q}\rightarrow 0}\frac{1}{ 2\beta}\sum_{i\omega_n}\int\frac{\mathrm{d^2}k}{(2\pi)^2}\left[G_1(i\omega_n,\vec{k})G_1(i\omega_n+i\nu_r,\vec{k}+\vec{q}) +g_2(i\omega_n,\vec{k})g_2(i\omega_n+i\nu_r,\vec{k}+\vec{q})\right]\nonumber\\
\frac{1}{2g}=\Lambda^{00}-\Lambda^{33}=-\lim_{i\nu_r\rightarrow 0,\vec{q}\rightarrow 0}\frac{1}{ 2\beta}\sum_{i\omega_n}\int\frac{\mathrm{d^2}k}{(2\pi)^2}\left[G_1(i\omega_n,\vec{k})g_2(i\omega_n+i\nu_r,\vec{k}+\vec{q}) +g_2(i\omega_n,\vec{k})G_1(i\omega_n+i\nu_r,\vec{k}+\vec{q})\right]\nonumber
\end{gather}

\ew

Taking the Matsubara sum we obtain 
\bea
2\Gamma&=&\sum_k\frac{V^2}{\left[(\eps^c_k-\eps^f_k)^2+4V^2\right]^{3/2}},\nonumber\\
\frac{1}{2g}&=&\sum_k\left[\frac{\sin^2\alpha_k f(-\eps_k)}{\eps_k-E_h}-\frac{\cos^2\alpha_k f(\eps_k)}{\eps_k-E_l}\right],
\eea
 where we have set the Fermi functions to $f(E^h)=1$ and $f(E^l)=0$, because we have a Kondo insulator. Carrying out the momentum integrals in two spatial dimensions we obtain in terms of the variables of appendix \ref{continuum}  \bea
2\Gamma&=&\frac{\nu}{4}\left[1+\frac{\eta}{\sqrt{1+\eta^2}}\right],\nonumber\\
\frac{1}{2g} &=&\nu\left[\frac{1}{2}-\eta^2+\eta\sqrt{1+\eta^2}\right]+\frac{\nu}{2}\frac{(\lambda+r_m\mu)^2}{V^2}.\qquad
\eea
These results can be simplified in the regime where $t_K={T_K}/{D}$ and $r_m={m_c}/{|m_f|}$ are both small and consequently $\eta\sim D/V\gg 1$ and ${\mu}/{D}\approx 2$:
 \bea
2\Gamma&\approx &\frac{\nu}{2}\\
\frac{1}{2g} &\approx&\nu{\cal Z}, \qquad {\cal Z}\equiv\left[1+t_K(1+r_m/t_K)^2\right].
\eea\\

\subsection{Terms quadratic in $A_x$}
Terms quadratic in gauge potential and containing only the spatial components are
\bea
\frac{\Delta F_{spatial}}{NT}&=&{\Lambda^{00}_{\nu\nu'}}\int{d\bar x_\mu}\sum_{a=0}^3A_\nu^a(\bar x_\mu)A_{\nu'}^{a}(\bar x_\mu)\nonumber\\
&&+{\Lambda^{33}_{\nu\nu'}}\int{d\bar x_\mu}\Big\{\sum_{a=0,3}-\sum_{a=1,2} \Big\}A^a_\nu(\bar x_\mu)A^a_{\nu'}(\bar x_\mu)\nonumber\\
&&+{4\Lambda^{03}_{\nu\nu'}}\int{d\bar x_\mu} A^0_\nu(\bar x_\mu)A^3_{\nu'}(\bar x_\mu)\nonumber\\
&&+\frac{\Lambda^{ff}_{\nu\nu'}}{2}\int{d\bar x_\mu}A^f_\nu (\bar x_\mu) A_{\nu'}^f(\bar x_\mu)\nonumber\\
&&+{\Lambda^{1f}_{\nu\nu'}}\int{d\bar x_\mu} [A_\nu^{0}(\bar x_\mu) +A_\nu^{3}(\bar x_\mu)]A_{\nu'}^f(\bar x_\mu)\nonumber
\eea
where we have defined
\bea
\Lambda^{ab}_{\nu\nu'}&\equiv&\lim_{i\nu_r\rightarrow 0,\vec{q}\rightarrow 0}\frac{1}{\beta}\sum_{i\omega_n}\int\frac{\mathrm{d^2}k}{(2\pi)^2}\frac{k_\nu}{m^a}G^a\left(i\omega_n-\frac{i\nu_r}{2},\vec{k}-\frac{\vec{q}}{2}\right)\nonumber\\
&&\hspace{1.8cm}\times\frac{k_{\nu'}}{m^b}G^b\left(i\omega_n+\frac{i\nu_r}{2},\vec{k}+\frac{\vec{q}}{2}\right).\qquad
\eea
This will turn out to be diagonal in lower indices $\Lambda^{ab}_{\nu\nu'}\propto\delta_{\nu\nu'}$. The structure of these terms is precisely equal to that of the $\Delta F_{2\tau}$ terms we studied before:
\bea
&&(\Lambda^{00}_{\nu\nu}+\Lambda^{33}_{\nu\nu})[(A^0_\nu)^2+(A_\nu^3)^2]+(\Lambda^
{00}_{\nu\nu}-\Lambda^{33}_{\nu\nu})[(A^1_\nu)^2+(A_\nu^2)^2]\nonumber\\
&&+4\Lambda^{03}_{\nu\nu}A_\nu^0A_\nu^3+\frac{1}{2}\Lambda^{ff}_{\nu\nu}(A_\nu^f)^2+\Lambda^{1f,1f}_{\nu\nu}[A_\nu^0+A_\nu^3]A_\nu^f,\qquad\label{eqc29}
\eea
but these have to be combined with the diamagnetic terms
\bea
\frac{\Delta F_{diag}}{NT}&=&\frac{2}{2m_c}\int{d\bar x_\mu}\Big\{\sum_{a=0}^3[A^a_\nu(\bar x_\mu)]^2G_C^{0}(0)\nonumber\\
&&\qquad\qquad +2A_\nu^0(\bar x_\mu)A_\nu^3(\bar x_\mu)G_C^3(0)\Big\}\nonumber\\
&&+\frac{1}{2m_f}\int{d\bar x_\mu}[A_\nu^f(\bar x_\mu)]^2G_f(0).
\eea
Computing the coefficients of \pref{eqc29} we have
\bw

\bean
\Lambda^{00}_{\nu\nu}+\Lambda^{33}_{\nu\nu}&=&\frac{1}{2\beta}\sum_{n,k}[G_1^2(i\omega_n,k)+g_2^2(i\omega_n,k)](v_\nu^c)^2= -\frac{1}{2\beta}\sum_{nk}G_{1f}^2(i\omega_n,k)v^c_\nu v^f_\nu -\frac{1}{2\beta}\sum_{nk}[G_1(i\omega_n,k)+g_2(i\omega_n,k)]\frac{1}{m_c},\\
\Lambda^{00}_{\nu\nu}-\Lambda^{33}_{\nu\nu}&=&\frac{1}{\beta}\sum_{nk}G_1(i\omega_n,k)g_2(i\omega_n,k)(v_\nu^c)^2,\\
4\Lambda^{03}_{\nu\nu}&=&\frac{1}{\beta}\sum_{nk}[G_1^2(i\omega_n,k)-g_2^2(i\omega_n,k)](v^c_\nu)^2=-\frac{1}{\beta}\sum_{nk}G_{1f}^2(i\omega_n,k)v_\nu^c v_\nu^f-\frac{1}{\beta}\sum_{nk}\sum_{nk}[G_1(i\omega_n,k)-g_2(i\omega_n,k)]\frac{1}{m_c},\\
\Lambda^{ff}_{\nu\nu}&=&\frac{1}{\beta}\sum_{nk}G_f^2(i\omega_n,k)(v_\nu^f)^2=-\frac{1}{\beta}\sum_{nk}G_{1f}^2(i\omega_n,k)v_\nu^cv_\nu^f+\frac{1}{\beta}\sum_{nk}G_f(i\omega_n,k)\frac{1}{|m_f|},\\
\Lambda^{1f,1f}_{\nu\nu}&=&\frac{1}{\beta}\sum_{nk}G_{1f}^2(i\omega_n,k)v_\nu^cv_\nu^f.
\eean
\ew
.Here, $v^c_\nu=k_\nu/m_c$, $v_\nu^f=-k_\nu/|m_f|$ and we have used $\partial_{k_{\nu}}G_1=G_1^2[v^c_\nu+V^2g_f^2v^f_\nu]$ and $G_{1f}=VG_1g_f$ to bring these terms to a form suitable to add the diamagnetic terms. The latter has the following coefficients:
\bean
\frac{1}{m_c}G_C^0(0)&=&\frac{1}{2\beta}\sum_{nk}[G_1(i\omega_n,k)+g_2(i\omega_n,k)]\frac{1}{m_c},\\
\frac{1}{m_c}G_C^3(0)&=&\frac{1}{2\beta}\sum_{nk}[G_1(i\omega_n,k)-g_2(i\omega_n,k)]\frac{1}{m_c},\\
-\frac{1}{2m_f}G_f(0)&=&\frac{1}{2\beta}\sum_{nk}G_f(i\omega_n,k)\frac{1}{m_f}.\\
\eean
These are exactly canceled by similar terms in $\Lambda$ coefficients, so that sum of the two has the form
\be
\frac{{\cal L}_{sp+dia}}{N}=\sum_{\nu=1}^d\Big\{2\Gamma v_{\Gamma}^2(A_\nu^0+A_\nu^3-A_\nu^f)^2+\frac{v_g^2}{2g}[(A_\nu^1)^2+(A_\nu^2)^2]\Big\},
\nonumber
\ee
with the coefficients
\bea
2\Gamma v_{\Gamma}^2&=&\frac{1}{2\beta}\sum_{nk}G_{1f}^2(i\omega_n,k)\frac{k_\nu^2}{m_cm_f}\\
\frac{v_g^2}{2g}&=&2\Gamma v_{\Gamma}^2 -\frac{1}{2\beta}\sum_{nk}[G_1(i\omega_n,k)-g_2(i\omega_n,k)]^2v_c^2,\qquad
\eea
again anticipating the Higgs- and the NL$\sigma$M terms. 
Note that in absence of magnetic coupling between the spins the $f$-electrons are localized $m_f\to\infty$ and consequently $ v_{\Gamma}^2 \to 0$. Carrying out the Matsubara sum, we obtain
\bw

\bea
2\Gamma v_{\Gamma}^2&=&\frac{1}{m_cm_f}\int\limits_{\mathrm{B.Z.}}\frac{\mathrm{d}^2k}{\left(2\pi\right)^2}\frac{V^2k_\nu^2}{\left[(\eps^c_k-\eps^f_k)^2+4V^2\right]^{3/2}},\\
\frac{v_g^2}{2g}&=&2\Gamma v_{\Gamma}^2 -\frac{1}{2m_c^2}\int\limits_{\mathrm{B.Z.}}\frac{\mathrm{d}^2k}{\left(2\pi\right)^2}\left\{ f'(\eps^c_k)-\frac{V^2}{\left[(\eps^c_k-\eps^f_k)^2+4V^2\right]^{3/2}}-\frac{4}{(\eps^c_k-\eps^f_k)+\sqrt{(\eps^c_k-\eps^f_k)^2+4V^2}}\right.\nonumber\\
&&\left.\qquad\qquad-\frac{2}{\sqrt{(\eps^c_k-\eps^f_k)^2+4V^2}}+\frac{2f(\eps^c_k)(\eps^c_k-\eps^f_k)}{V^2}\right\}k_\nu^2,
\eea

\ew
 where we have again set the Fermi functions to $f(E^h)=1$ and $f(E^l)=0$, because we have a Kondo insulator at $T \ll T_K$. In this limit the momentum integrals can be obtained analytically for the continuum model in $d=2$ spatial dimensions:

\bw

\bea
2\Gamma v_{\Gamma}^2&=&\frac{\pi\nu^2 V}{2m_cm_f}\left[\sqrt{1+\eta^2}+\eta\right]\\
\frac{v_g^2}{2g}&=&2\left(1+\frac{m_f}{m_c}\right)\Gamma v_{\Gamma}^2-\frac{1}{2\pi}\left\{\eta-\frac{4}{3}\left[(\eta^2+1)^{3/2}+\eta^3\right]+2\sqrt{\eta^2+1}+2\eta^2\left[\eta+\sqrt{\eta^2+1}\right]\right\}+\frac{\mu}{4\pi}+\frac{\mu^3}{12\pi V^2}\qquad\eea
\ew

To lowest order in ${T_K}/{D}$ and $r_m={m_c}/{m_f}$ these coefficients simplify to

\bea
2\Gamma v_{\Gamma}^2&\approx&\frac{\pi}{2}r_mD,\\
\frac{v_g^2}{2g}&\approx& \frac{D}{\pi}.
\eea

\subsection{Alternative representation} \label{altrep}
Using the parametrization \pref{eqgparam} we have
\be
(A^1_\nu)^2+(A^2_\nu)^2=(\partial_\nu\theta)^2+(\partial_\nu\phi)^2\sin^2\theta=(\partial_\nu\vec n)^2
\ee
where we have used the Hopf map $\vec n=z\dg\vec\sigma z$. Moreover,
\be
A^3_\nu=\partial_\nu\varphi+\partial_\nu\phi\cos\theta=-iz\dg\partial_\nu z
\ee
Note that the `magnetic field' associated with this vector potential is equal to the topological charge
\be
B_\mu=\eps_{\mu\lambda\nu}\partial_\lambda A^3_\nu=\eps_{\mu\lambda\nu}\vec n\cdot(\partial_\lambda\vec n\times \partial_\nu \vec n).
\ee
For the example, in two dimensions, $B_z$ is the density of the skyrmions.

The coefficients computed in the previous section can be used to write the Lagrangian in the following form
\bea
{\cal L}&=&\frac{1}{2g}\Big[(\partial_\tau \vec n)^2+v_g^2\sum_{a=1}^d(\partial_a\vec n)^2\Big]\\
&&+\frac{\Gamma}{2}\Big[
({\textstyle \frac{1}{2}}\Omega^{z}_{\tau }-(A^{ext}_\tau-A^f_\tau))^2
+v_\Gamma^2\sum_{a=1}^d
({\textstyle \frac{1}{2}}\Omega^{z}_{a}-(A^{ext}_a-A^f_a))^2
\Big],\nonumber
\eea
where we have restored $A^{3}_{\mu}= {\frac{1}{2}}\Omega^{z}_{\mu}$
and 
the parameters are given by
\bea
\frac{1}{2g}&\approx&\nu{\cal Z}\\
2\Gamma&\approx&\frac{\nu}{2}\\
v^2_g&\approx&\frac{D}{\pi\nu{\cal Z}}\\
v_\Gamma^2&\approx&\frac{\pi r_mD}{2\nu}.
\eea
where ${\cal Z}=\left[1+t_K(1+r_m/t_K)^2\right]$  was defined before. Using $D=\tilde q_f/\nu$, $\nu\approx \rho=m_c/2\pi$ and $\tilde q_f=k_F^2/4\pi$ we find the expressions reported in Eq.\,\pref{eqcoefmain}.
\bibliography{2ck}
\end{document}